\documentclass[aps,twocolumn,showpacs,nofootinbib,showkeys,superscriptaddress,preprintnumbers,longbibliography,10pt]{revtex4-2}
\usepackage[utf8]{inputenc}

\usepackage{latexsym}
\usepackage{epsfig}
\usepackage{graphicx,subfigure}
\usepackage[normalem]{ulem}
\usepackage{color}
\usepackage{xcolor}
\usepackage{multirow}
\usepackage{hyperref}
\usepackage{numprint}
\usepackage{eucal}
\usepackage{amsmath}
\usepackage{amssymb}
\usepackage{orcidlink}
\usepackage{xr}

\begin{document}

\def\ga{\mathrel{\raise.3ex\hbox{$>$\kern-.75em\lower1ex\hbox{$\sim$}}}}
\def\la{\mathrel{\raise.3ex\hbox{$<$\kern-.75em\lower1ex\hbox{$\sim$}}}}

\def\be{\begin{equation}}
\def\ee{\end{equation}}
\def\bea{\begin{eqnarray}}
\def\eea{\end{eqnarray}}

\def\betap{\tilde\beta}
\def\del{\delta_{\rm PBH}^{\rm local}}
\def\Msun{M_\odot}

\newcommand{\dd}{\mathrm{d}} 
\newcommand{\Mpl}{M_P} 
\newcommand{\mpl}{m_\mathrm{pl}} 

\newcommand{\IUCAA}{Inter-University Centre for Astronomy and
  Astrophysics, Post Bag 4, Ganeshkhind, Pune 411 007, India}

\newcommand{\UmasD}{Department of Physics, University of Massachusetts, Dartmouth, MA 02747, The USA}

\newcommand{\NIKHEF}{Nikhef - National Institute for Subatomic Physics, Science Park, 1098 XG Amsterdam, The Netherlands}

\newcommand{\UVA}{Institute for High-Energy Physics, University of Amsterdam, Science Park, 1098 XG Amsterdam, The Netherlands}
 
\newcommand{\GRASP}{Institute for Gravitational and Subatomic Physics, Utrecht University, Princetonplein 1, 3584 CC Utrecht, The Netherlands}

\newcommand{\LIEGE}{Department of Astrophysics, Geophysics and Oceanography (GEO), Space sciences, Technologies and Astrophysics Research (STAR), Universit\'e de Li\`ege, allée du six Août 19,
4000 Liège, Belgium}

\newcommand{\ULB}{Service de Physique Th\'eorique, Universit\'e Libre de Bruxelles (ULB), Boulevard du Triomphe, CP225, B-1050 Brussels, Belgium}

\newcommand{\LOUVAIN}{Center of Cosmology, Phenomenology and Particle Physics, Research Institute in Mathematics and Physics (IRMP), Louvain University, 2 Chemin du Cyclotron, 1348 Louvain-la-Neuve, Belgium}

\newcommand{\ICTS}{International Centre for Theoretical Sciences, Tata Institute of Fundamental Research, Bangalore 560089, India}

\newcommand{\CALTECH}{LIGO, California Institute of Technology, Pasadena, CA 91125, USA}
 
 \newcommand{\RESCEU}{Research Center for the Early Universe (RESCEU), University of Tokyo, Tokyo, 113-0033, Japan}
 
 \newcommand{\UoB}{School of Physics and Astronomy and Institute for Gravitational Wave Astronomy,\\University of Birmingham, Edgbaston, Birmingham, B15 9TT, United Kingdom}
 
\thispagestyle{empty}

\title{Search for sub-solar primordial black holes in low mass ratio binaries with LIGO-Virgo O2 data and implications for the primordial black hole dark matter fraction} 

\author{Khun Sang Phukon\orcidlink{0000-0003-1561-0760}}\email{k.s.phukon@bham.ac.uk}
\affiliation{\NIKHEF}
\affiliation{\UVA}
\affiliation{\GRASP}
\affiliation{\UoB}

\author{Gregory Baltus}
\affiliation{\LIEGE}

\author{Sarah Caudill}
\affiliation{\GRASP}
\affiliation{\NIKHEF}
\affiliation{\UmasD}

\author{Sebastien Clesse}\affiliation{\ULB}\email{sebastien.clesse@ulb.be}

\author{Antoine Depasse}
\affiliation{\LOUVAIN}

\author{Maxime Fays}
\affiliation{\LIEGE}

\author{Heather Fong}
\affiliation{\RESCEU}
\affiliation{University of British Columbia, Vancouver Campus,  2329 West Mall, Vancouver, British Columbia V6T 1Z4, Canada}

\author{Shasvath J. Kapadia}
\affiliation{\ICTS}

\affiliation{\IUCAA}

\author{Ryan Magee}
\affiliation{\CALTECH}

\author{Andres Jorge Tanasijczuk}
\affiliation{\LOUVAIN}

\begin{abstract}  
We perform a search for binary black hole mergers with one sub-solar mass (SSM) black hole and a primary component above $\sim 2 M_\odot$ in data from the second observing run (O2) of the LIGO-Virgo detectors. Our analysis extends the parameter space explored by previous LIGO-Virgo Collaboration searches for binaries containing SSM components into a region of parameter space motivated by broad mass distributions of primordial black holes (PBHs) exhibiting a peak around $[2-3] M^{}_\odot$, which can arise from the reduction of the equation of state during the QCD phase transition in the early Universe. Four candidate events are found passing a signal-to-noise ratio (SNR) threshold of 8 and a false alarm rate (FAR) threshold of 2 per year, although none are statistically significant enough to constitute a confident detection. Assuming a null result for the search, we derive PBH model-independent 90\% confidence upper limits on the PBH merger rates by estimating the sensitive volume-time of the search using simulated gravitational-wave signal injections. We interpret these observational limits using a representative broad PBH mass function bearing imprints of the thermal history of the early Universe and considering both early and late PBH binary formation channels. The resulting constraints on the PBH dark-matter fraction, $f^{}_{\rm PBH}$, depend on the assumed mass function and merger-rate prescription.  For all considerations, the upper limits remain above $f^{}_{\rm PBH}=1$, indicating the O2 data from LIGO-Virgo are not sensitive enough to place for meaningful constrain within the assumptions of the PBH mass model and PBH binary mergers.

\end{abstract}

\maketitle

\section{Introduction}

The first detection of a gravitational wave (GW) event (GW150914)~\cite{Abbott:2016blz} has ushered astronomy into a GW era.  Since then a large population of binary black hole (BBH) mergers~\cite{Abbott:2020tfl, Abbott:2020khf, LIGOScientific:2020stg, Abbott:2017oio, Abbott:2016nmj, Abbott:2017gyy, Abbott:2017vtc, LIGOScientific:2018mvr, Abbott:2020niy,TheLIGOScientific:2016pea,LIGOScientific:2026wfs,LIGOScientific:2026ctl} have been observed by Advanced LIGO~\cite{TheLIGOScientific:2014jea}, Advanced Virgo~\cite{TheVirgo:2014hva} and KAGRA~\cite{Aso:2013eba,KAGRA:2020tym,Somiya:2011np}, revealing some intriguing and mostly unexpected properties of BBHs:  large masses, low spins, the existence of black holes (BHs) in the low mass gap~\cite{Abbott:2020khf} or in the pair-instability mass gap~\cite{Abbott:2020tfl} and with low mass ratios~\cite{Abbott:2020khf}.  Elucidating the origin of LIGO-Virgo BHs has emerged as an important research topic. One possible interpretation is that a fraction of the observed black holes (BHs) may be primordial in origin~\cite{Bird:2016dcv, Clesse:2016vqa, Sasaki:2016jop}. 

Primordial black holes may have formed in the early Universe due to the gravitational collapse of large overdensities~\cite{Zeldovich:1967lct,Hawking:1971ei,Carr:1974nx,1975Natur.253..251C}, e.g.~coming from inflation~\cite{Carr:1993aq,Ivanov:1994pa,GarciaBellido:1996qt,Kim:1996hr}, and could explain from a fraction to the totality of the dark matter (DM).  There exist a series of astrophysical and cosmological limits on their abundance, covering almost all possible masses (see e.g.\cite{Carr:2016drx,Carr:2009jm} for reviews on PBHs and~\cite{Carr:2020gox,Carr:2020xqk,Green:2020jor} for recent developments), as well as possible clues in observations~\cite{clesse2018seven,Carr:2019kxo}.  Those limits and observational signatures are PBH model dependent, and the viability of PBHs as a substantial component of dark matter and possible explanations of all GW events remains debated. 

Contrary to stellar BHs, there is no physical process preventing the formation of PBHs lighter than the Chandrasekhar mass~\cite{chandra} or in the pair-instability mass gap~\cite{Rakavy_Shaviv_67,1968Ap&SS...2...96F}.  The sub-solar and intermediate mass ranges are thus ideal targets to distinguish between stellar and primordial origins.  If some of the observed BH mergers are primordial, one can expect a relatively broad mass distribution of PBHs spreading within these interesting regions. The thermal history of the Universe at the QCD epoch and the resulting transient reduction of the equation-of-state  could lead to such features in the PBH mass distribution~\cite{Niemeyer:1997mt,Jedamzik:1996mr,Byrnes:2018clq}, in particular, a peak around the solar-mass and a bump around $30 M_\odot$~\cite{Byrnes:2018clq}. Such mass functions have been proposed as possible explanations for some LIGO-Virgo BBH mergers\cite{Carr:2019kxo,Clesse:2020ghq,Jedamzik:2020omx,Jedamzik:2020ypm} and for other astrophysical phenomena that may provide indirect evidence for BH populations, including selected OGLE microlensing events indicating low mass BH~\cite{2020A&A...636A..20W}, and correlations between the infrared and X-ray backgrounds~\cite{Kashlinsky:2016sdv,2018RvMP...90b5006K,2005Natur.438...45K}. However, the PBH interpretation of these observations remain model dependent. All this provides strong theoretical and observational motivations to extend previous SSM object searches in the first~\cite{Abbott:2018oah} and second~\cite{LIGOScientific:2019kan} observing runs of LIGO-Virgo (referred as O1 and O2) and to search for binaries combining an SSM black hole and a primary component above $\sim 2 M_\odot$, potentially coming from the QCD peak.

In this paper, we perform an extended search on O2 data, 
assuming one SSM component between $0.19 \mathrm{M}_\odot$ and $1M_\odot$ and a primary component between $1.95$ and $11 \mathrm{M}_\odot$.  Our analysis is therefore complementary to previous searches, either restricted to $2 \mathrm{M}_\odot$ for the primary component~\cite{Abbott:2018oah,LIGOScientific:2019kan} or considering even lower mass ratios~\cite{Nitz:2020bdb}.  We derive new model-independent merger rate limits in the $(m_1,m_2)$ plane of the two component masses\footnote{When our work was almost completed, an independent search has been released with a similar mass range~\cite{Nitz:2021mzz}.  Besides providing an independent confirmation of their results, our analysis also reveals a few candidate events and includes new limits for some motivated realizations of the PBH mass function. The asymmetric mass parameter space of this work is also included in subsequent SSM object searches by the LIGO-Virgo-KAGRA Collaboration~\cite{LIGOScientific:2021job,LIGOScientific:2022hai,LIGOScientific:2026wxz}, which came online after submission of this paper. }.  Furthermore, whereas previous limits on $f^{}_{\rm PBH}$ from O1 and O2 data~\cite{Abbott:2018oah,LIGOScientific:2019kan}-- defined as the PBH abundance with respect to DM -- relied on the merging rates of early PBH binaries from~\cite{Sasaki:2016jop}, we include the rate suppression due to binary disruption by early forming clusters, matter inhomogeneities and nearby PBHs, put in evidence with N-body simulations~\cite{Raidal:2018bbj}. We consider in addition, the merging rate of late PBH binaries formed in clusters.  

The paper is organized as follows. In Sec.~\ref{sec:search}, we describe the search analysis and present search results, and derive  model-independent constraints on upper limit on the merger rate of binaries involving one SSM object. Section~\ref{sec:pbh_models} introduces the PBH mass  model considered in this paper and merger rates associated with formation channels of PBH binaries.  The constraints on the PBH dark matter fraction model for the PBH binary models are presented in Sec.~\ref{sec:pbh_constraints}. We summarize and conclude in Sec.~\ref{sec:conclusion}.

\section{Search, candidates and rate limits}
\label{sec:search}
We  analyze  O2  public  data  from single and coincident observation time of the two LIGO detectors, comprising $\sim$193 days of analysis time.   This search uses the GstLAL-based inspiral pipeline~\cite{Cannon:2020qnf, cannon2012toward, Hanna:2019ezx, messick2017analysis, sachdev2019gstlal, lalsuite} with configurations and procedures as outlined in~\cite{Abbott:2020niy}.
Data from each detector are matched-filtered~\cite{cannon2012toward,Dhurandhar:1992mw,Balasubramanian:1995bm,Owen:1995tm,Owen:1998dk,Allen:2005fk} with an (anti-)aligned low-spin template bank of \numprint{195468} template waveforms modelled with the frequency-domain waveform-approximant, TaylorF2~\cite{Blanchet:1995ez,Poisson:1997ha, Damour:2001bu, Damour:2000zb,Buonanno:2009zt,Mikoczi:2005dn,Blanchet:2005tk,Arun:2008kb,Bohe:2013cla,Bohe:2015ana,Mishra:2016whh}. Components have primary mass $m_1 \in [1.95, 11.0 ] \mathrm{M}_\odot$, SSM $m_2 \in [0.19, 1.0] \mathrm{M}_\odot$ and spin magnitude $\chi_{i,z} \in [0, 0.1 ]$.
The total masses ($\mathrm{M}_{\mathrm{tot}}$) and mass ratios ($q = m_2/m_1$) of the search are limited between $\mathrm{M}_{\mathrm{tot}} \in  [{2.2 -  11.0}]\, \mathrm{M}_\odot$ and $q \geq 0.1$, respectively.  The template bank construction uses the stochastic method of Refs.~\cite{Harry:2008yn,Babak:2008rb, Harry:2009ea, Manca:2009xw} with a  minimal match of $0.97$. Following Ref.~\cite{Magee:2018opb}, we perform matched-filtering of the data in the frequency band 45-1024 Hz in order to reduce the computational burden of filtering data.

Candidates are ranked using a likelihood-ratio $\mathcal{L}$ ranking statistic~\cite{Cannon:2015gha,messick2017analysis,Hanna:2019ezx}. The ranking statistics incorporates a log-uniform mass model for signals in the likelihood-ratio construction~\cite{Fong2018Thesis}. Here we note, in contrast to Ref.~\cite{Abbott:2020niy}, our pipeline does not include data-quality penalties to rank single-detector candidates~\cite{Essick:2020qpo,Godwin:2020weu}. This ranking statistic also provides an algebraic procedure to estimate the significance of each event~\cite{Cannon:2012zt, messick2017analysis}.  The significance of an event is expressed in terms of a false alarm rate (FAR), e.g. defined in Section IV.C of Ref.~\cite{messick2017analysis}.

\begin{figure}[t]
\begin{center}
\includegraphics[width = 0.47 \textwidth]{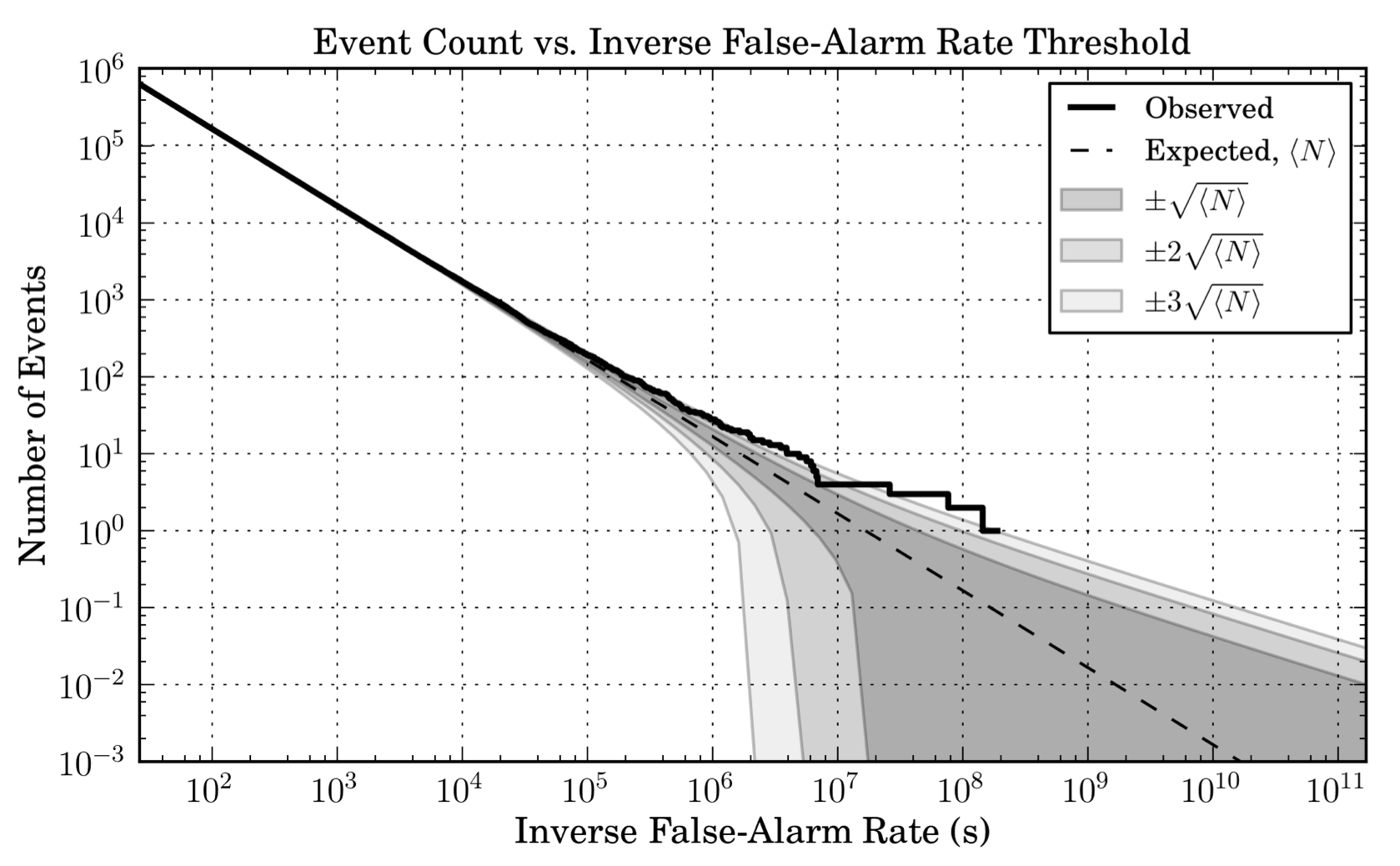}
\caption{\label{Search_summary} Results of the sub-solar PBH search in O2, in terms of number of events as a function of their inverse FAR.  The dashed line is the expected distribution of background triggers, with the gray bands indicating uncertainties in multiples of the standard deviation for a Poisson distribution.  The four candidates reported in Table~\ref{tab:candidate} lie slightly above the $3\sigma$ limit.
}
\label{fig:search_summary}
\end{center}
\end{figure}

Search results are presented in Fig.~\ref{Search_summary}. Table~\ref{tab:candidate} reports the top four candidates with a signal-to-noise ratio SNR$>8$ and a FAR $<2 \, {\rm yr}^{-1}$.  Time-frequency representations of the whitened data around each candidate are provided in Fig.~\ref{fig:omegascans} in the Supplementary Material for different resolutions and do not reveal any clear GW signature.  The first two candidates are single detector triggers, implying their lower FAR values may have more uncertainty. The remaining candidates have high FARs considering the number of days analyzed and in comparison to GW events reported in~\cite{LIGOScientific:2018mvr,Abbott:2020niy}. With these considerations, it is not clear if any of the four candidates are real GW events. However, a subsequent Bayesian parameter estimation study of third candidate event from 2017-04-01 in~\cite{Morras:2023jvb} including a subset of the authors of this work found that the posterior distribution favors a secondary mass $m_2<1\,M^{}_\odot$ at an 84\% credible level. Although this evidence is not sufficient to establish the signal nature of the candidate at search level. For the rest of the paper, we assume a null result of the search.

\begin{table*}[ht]
\caption{The candidates of the search with a SNR $>8$ and a FAR $<2 \,{\rm yr}^{-1}$. 
Here we report the FAR, $\ln \mathcal{L}$, the UTC time of the event (date and hours), and recovered template parameters.}
\begin{ruledtabular}
\begin{tabular}{c c c c c c c c c c}
FAR [yr$^{-1}$] &  $\ln\mathcal{L}$ & UTC time & mass 1 $[M_{\odot}]$ & mass 2 $[M_{\odot}]$ & spin1z &  spin2z & Network SNR & H1 SNR & L1 SNR  \\
\hline\\
0.1674 & 8.457 & 2017-03-15 15:51:30 & 3.062 & 0.9281 & 0.08254 & -0.09841 & 8.527 & 8.527 & - \\
0.2193  & 8.2 & 2017-07-10 17:52:43 & 2.106 & 0.2759 & 0.08703	&0.0753	 & 8.157 & - & 8.157 \\
0.4134 & 7.585 & 2017-04-01 01:43:34 & 4.897 & 0.7795 &-0.05488	 &-0.04856	 & 8.672 & 6.319 & 5.939 \\
1.2148 & 6.589 & 2017-03-08 07:07:18 & 2.257 & 0.6997 &-0.03655	 & -0.04473	& 8.535 & 6.321 & 5.736 \\
\end{tabular}
\end{ruledtabular}
\label{tab:candidate}
\end{table*}

\begin{figure}[t]
\centering
\includegraphics[width =0.49\textwidth]{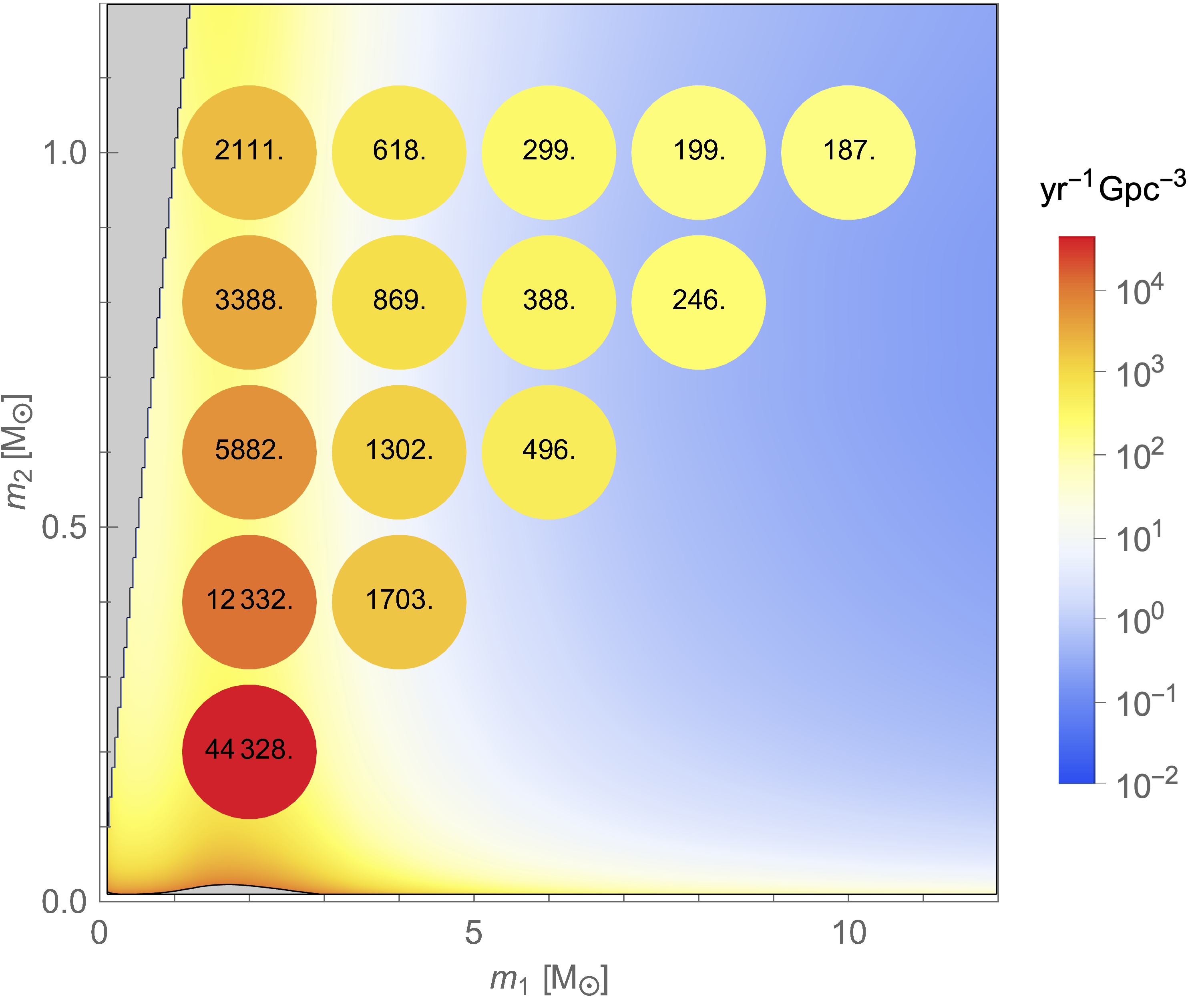} 
\caption{\label{fig:2Dlimit}
The disk color and numbers inside represent the merger rate limits 
at $90\%$ CL $(\mathcal{R}_{90\%})$, for component masses between $2$ and $10\, M_\odot$ for the primary component and between $0.2$ and $1 \, M_\odot$ for the secondary component, with mass ratios $q > 0.1$, in respective bin sizes of $2$ and $0.2 \, M_\odot$.   The color scale behind represents the predictions of the EB merging rates for our mass model~\cite{Carr:2019kxo}, assuming $n_{\rm s} = 0.97$, $\gamma = 0.8$ and $f_{\rm PBH} = 1$.} 
\end{figure}

Following methods in Refs.~\cite{Farr:2013yna, Kapadia:2019uut}, we place $90\%$ confidence level (CL) upper limits on the rate of mergers, $\mathcal{R}_{90\%}$, of binaries with an SSM component without invoking any PBH model dependence for the discrete mass pairs shown in Fig.~\ref{fig:2Dlimit} and Table~\ref{space-time-rate-table}. We first estimate the sensitive space-time volume $\langle \mathrm{VT} \rangle$ of our search by using 15 populations of simulated, non-spinning binary sources, spanning the $(m_1, m_2)$ plane of our search. Each simulated population represents binaries with fixed intrinsic masses, given in Table~\ref{space-time-rate-table} and their distributions are isotropic in sky locations and source orientations, and uniform in co-moving distances.  We simulate $\mathcal{O}(10^6)$ sources  using the time-domain \texttt{TaylorT4} waveforms with phase corrections at 3.5 PN order~\cite{Buonanno:2009zt} and inject them in data during the observation period $T_{\rm obs}$ of O2.  We search for all these injected signals to measure $\langle \mathrm{VT} \rangle$ for each population.  Subsequently, we use the estimated $\langle \mathrm{VT} \rangle$ to calculate upper limits on the merger rate $\mathcal{R}_{90\%}$. The values of $\langle\mathrm{VT}\rangle$ and the corresponding values of $\mathcal{R}_{90\%}$ for all simulated populations are provided in Table~\ref{space-time-rate-table}. These merger rate limits constitute the model-independent observational result of this search. Their interpretation in terms of the PBH abundance is presented separately in Sec.~\ref{sec:pbh_models} using representative PBH mass function and merger rate models.

\begin{table}[t]
\centering
\caption{\label{space-time-rate-table} Sensitive volume-time $\langle \mathrm{VT} \rangle$ and PBH model-independent merger rate upper limit at 90\%  $\mathcal{R}_{90\%}$. The first column shows the fixed component masses of binaries of a given injection population.}
\begin{ruledtabular}
 \begin{tabular}{l c c }
         $(m1, m2)\, M_\odot$ & $\langle \mathrm{VT} \rangle$ ($\mathrm{Gpc}^3 \text{yr} $) & $ \mathcal{R}_{90\%}\, (\text{Gpc}^{-3} \mathrm{yr}^{-1})$  \\
        \hline\\
        2.0, 0.21 &    $6.06 \times 10^{-5}$ &   45371\\
        2.0, 0.40 &    $2.18 \times 10^{-4}$ &   12615\\
        2.0, 0.60 &   $4.57 \times 10^{-4}$ &   6018\\ 
        2.0, 0.80 &    $7.94 \times 10^{-4}$&   3467\\ 
        2.0, 0.99 &     $1.27 \times 10^{-3}$&  2159\\ 
        4.0, 0.41 &    $1.57 \times 10^{-3}$&   1743\\ 
        4.0, 0.60 &   $2.06 \times 10^{-3}$ &  1332\\ 
        4.0, 0.80 &    $3.09 \times 10^{-3}$ &   889\\ 
        4.0, 0.99 &    $4.35 \times 10^{-3}$ &   633\\ 
        6.0, 0.61 &    $5.42 \times 10^{-3}$ &   507\\ 
        6.0, 0.80 &    $6.92 \times 10^{-3}$ &   398\\ 
        6.0, 0.99 &    $8.99 \times 10^{-3}$ &   306\\ 
        8.0, 0.81 &   $1.09 \times 10^{-2}$ &   252\\ 
        8.0, 0.99 &    $1.34 \times 10^{-2}$ &   204\\ 
        9.8, 0.99 &   $1.43 \times 10^{-2}$  &  192\\ 
    \end{tabular}
    \end{ruledtabular}
\end{table}

\section{PBH mass and rate distribution}  
\label{sec:pbh_models}

The recent wide mass distribution $f(m^{}_{\rm PBH})$  proposed by Carr {\it et al.} in~\cite{Carr:2019kxo} has been considered in our analysis.  It includes the expected features from the QCD transition and relies on an almost scale invariant primordial power spectrum of density fluctuations, enhanced compared to cosmological scales, with a spectral index $n_{\rm s} = 0.97$ (values of $n_{\rm s}$ between $0.96$ and $0.98$ are realistic). The PBH peak is located at the solar-mass scale and there is no overproduction of PBHs at smaller and larger masses. It was argued that, even if $f^{}_{\rm PBH} = 1$ the model can evade the microlensing limits derived assuming a smooth PBH distribution because of PBHs clusters seeded by the Poisson fluctuations in the PBH distribution at formation~\cite{Clesse:2020ghq}. However, this argument does not hold as recently it has been shown that PBH clustering can not weaken  mircrolensing constraints on $f^{}_{\rm PBH}$, thereby preventing PBH to be the entirety of the DM in stellar mass region~\cite{Gorton:2022fyb,Petac:2022rio, DeLuca:2022uvz}. Assuming an almost scale invariant primordial power spectrum allows us to set relatively conservative limits, because peaked spectra or with a negative running of the spectral index typically lead a lower number density of PBHs both at high and low mass. We consider two plausible values of the ratio between the PBH and the Hubble horizon masses at formation, $\gamma = 0.8$ and $\gamma = 0.2$ for the normalized mass density distribution $f(m^{}_{\rm PBH})$. The first value leads to a peak around $2.5 M_\odot$ motivated by GW190425~\cite{Abbott:2020uma} and GW190814~\cite{Abbott:2020khf}.  
The second one is obtained by considering the turnaround scale~\cite{Byrnes:2018clq} in the PBH gravitational collapse. PBH binaries can form through two channels whose associated merging rates are briefly introduced below.

\textbf{Merging rate of Early Binaries (EB).} PBH binaries are formed in the early Universe, between the PBH formation time and the matter-radiation equality.  A fraction of them will merge nowadays with a merging rate distribution (per unit logarithmic mass of the two black holes) given by~\cite{Raidal:2018bbj,Ali-Haimoud:2017rtz,Gow:2019pok,Liu:2019rnx,Kocsis_2018,Hutsi:2020sol,Clesse:2020ghq}
\bea
        R_{\rm EB} &= & 
        \frac{1.6 \times 10^6}{\rm Gpc^3 yr} f_{\rm sup} f_{\rm PBH}^{34/37} \left(\frac{m_1 + m_2}{M_\odot}\right)^{-32/37} \nonumber \\ & \times & \left[\frac{m_1 m_2}{(m_1+m_2)^2}\right]^{-34/37} f(m_1) f(m_2)~,  \label{eq:cosmomerg}
\eea
where $m_1$ and $m_2$ are the two binary component masses.  N-body simulations have shown that the original merging rates from~\cite{Sasaki:2016jop} are somehow suppressed~\cite{Raidal:2018bbj}, which motivates the introduction of a suppression factor $f_{\rm sup} $ plausibly ranging between $10^{-3}$ and $0.1$. Our assumption for $f_{\rm sup} $ in relation with previous works of~\cite{Raidal:2018bbj,Hutsi:2020sol} is described in the Supplementary Material. As detailed in the Sec. III of the Supplementary Material, neglecting this suppression factor results in constraints on $f_{\rm PBH}$ that are approximately two orders of magnitude tighter, demonstrating that its inclusion is critical for obtaining a conservative bounds on $f_{\rm PBH}$. We do not consider the merging rate of disrupted binaries, estimated in~\cite{Vaskonen:2019jpv} for a monochromatic mass function and which may surpass the EB rate if $f_{\rm PBH}$ is close to one, because it is unclear how to extend these estimations for wide mass distributions like the ones we consider here.

\textbf{Merging rate of Late Binaries (LB).}   PBH binaries also form by tidal capture in PBH clusters, which gives the following rate distribution \cite{Clesse:2020ghq}, 
\be
        R_{\rm LB} \equiv  
        R_{\rm clust.}  f_{\rm PBH} f(m_1) f(m_2) 
        \frac{(m_1 + m_2)^{10/7}}{(m_1 m_2)^{5/7}},  \label{eq:cosmomergcapt}
\ee
where $R_{\rm clust.} $ is an effective scaling factor that incorporates the PBH clustering  properties.  We consider $R_{\rm clust} \approx 420\, {\rm yr}^{-1}{\rm Gpc}^{-3}$ as a benchmark,  following~\cite{Clesse:2020ghq}, and comprised within $[80-1770]~{\rm yr}^{-1}{\rm Gpc}^{-3}$ for our estimations of uncertainty bands\footnote{This range for $R_{\rm clust}$ is obtained when imposing the 90\% CL rate limits from GW190425 at $m_1 = 2.6 M_\odot$ and $m_2 = 2.0 M_\odot$ with $\gamma = 0.8$ in the mass model from~\cite{Carr:2019kxo}.}.  This value is larger than one may expect from standard halo mass functions~\cite{Bird:2016dcv} but is realistic if one takes into account the additional clustering due to Poisson fluctuations in the initial PBH separation and the relaxation time of small DM halos. As shown in~\cite{Clesse:2020ghq}, with such considerations, LBs compete with the rate of EBs and could produce the rates inferred from the GW events GW190425~\cite{LIGOScientific:2020aai}, GW190814~\cite{LIGOScientific:2020zkf} and GW190521~\cite{LIGOScientific:2020iuh}.

\begin{figure}[t]
\centering
\includegraphics[width = 0.49 \textwidth]{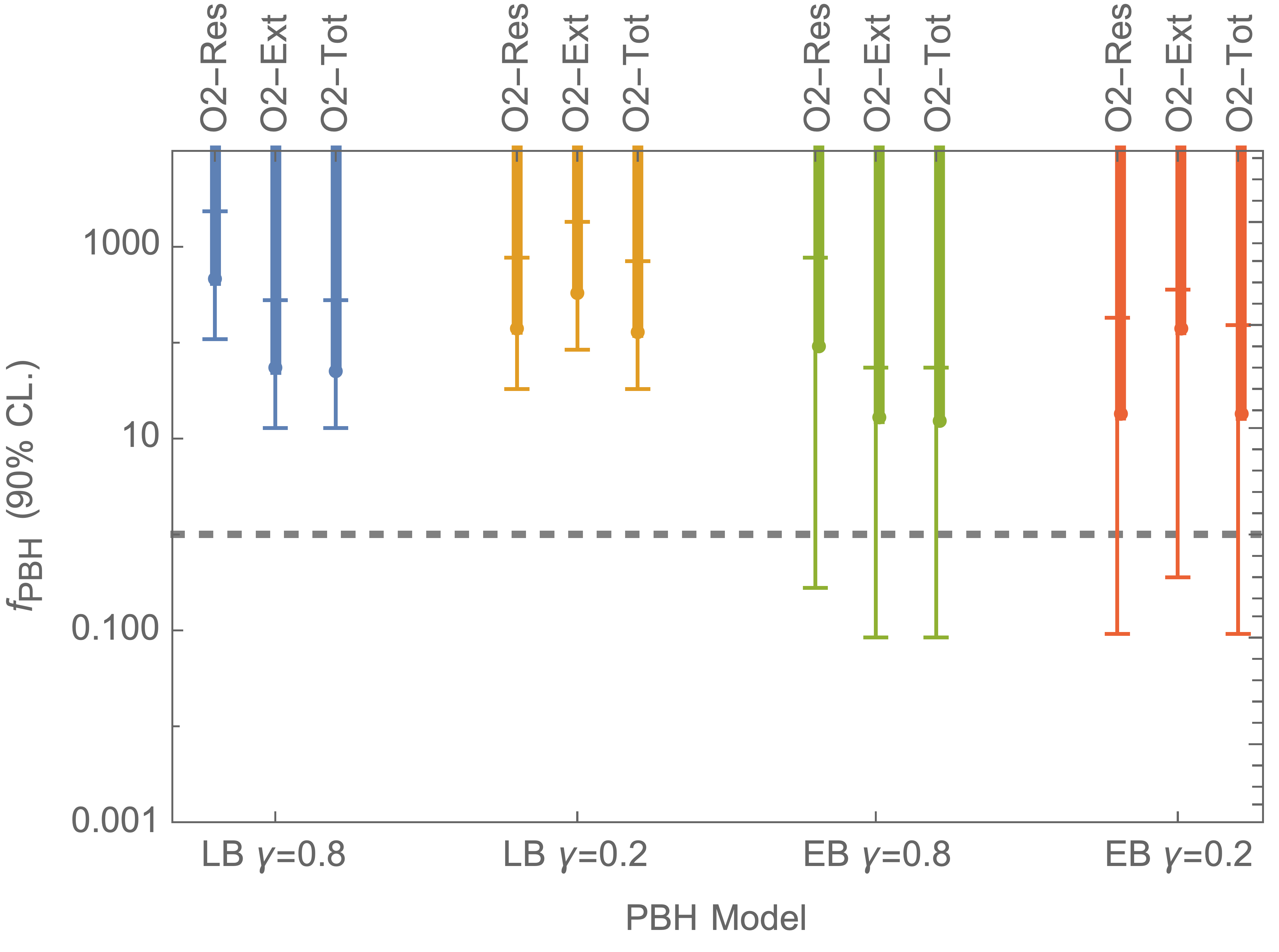}
\caption{\label{fig:fPBHlimit} Limits (90\% CL) on $f_{\rm PBH}$ for the broad PBH mass function of Ref.~\cite{Carr:2019kxo}, for late (LB) and early (EB) binaries for $\gamma = 0.8$ and $0.2$, from the search results in O2 data restricted to $m_1 < 2 M_\odot$~\cite{LIGOScientific:2019kan} (O2-Res), this search (O2-Ext) and their combination (O2-Tot).  The error bars estimate the uncertainties from the rate parameters $R_{\rm clust}$ and $f_{\rm sup}$ as described in the Supplementary Material.  The dashed horizontal lines denotes $f^{}_{\rm PBH} = 1$; limits above this line indicate that the observations
do not constrain $f^{}_{\rm PBH}$ within the physical range.}
\end{figure}

\section{Constraints on PBH models}
\label{sec:pbh_constraints}
The EB merging rate predictions for mass bin widths $\Delta m_1 = 2 M_\odot$ and $\Delta m_2 = 0.2 M_\odot$, $f_{\rm PBH} = 1$ and $\gamma = 0.8$ have been represented in Fig.~\ref{fig:2Dlimit} with the corresponding rate limits.  The other cases typically lead to lower rates in the mass region probed by the search.  A similar figure for LBs is included in the Supplementary Material, where we also compare the rates for $m_1 < 2M_\odot$ to the limits from \cite{Abbott:2018oah,LIGOScientific:2019kan}.
In all cases, we find that the subsolar limits do not yet exclude that $f^{}_{\rm PBH} = 1$ in O2 data.  Therefore, more sensitive data will be needed to probe these broad-mass PBH models. For EBs the rate predictions are quite close to the current limits.  If some candidates were to be real GW events, they could be explained by EBs given the uncertainties on the rate suppression parameter $f_{\rm sup}$.  But it is less likely that they come from LBs as EB rates are higher than LB rates in the parameter space considered here~\cite{Raidal:2024bmm}.  A firm detection would therefore help to distinguish the possible mass distributions and binary formation channels.  For each case, we have computed an upper limit on $f^{}_{\rm PBH}$, shown in Fig.~\ref{fig:fPBHlimit} with error bars corresponding to the estimated rate uncertainties. For this purpose, we assume a Poissonian distribution of GW events and build a simple $\chi^2$ function for $f_{\rm PBH}$,
\bea
\chi^2 & = & \sum_{{\rm bins\, } i,j} \left[ \frac{R_{\rm EB/LB} (f_{\rm PBH},m_{1,i},m_{2,j}) \Delta m_1 \Delta m_2 }{ m_{1,i} m_{2,j} \times \mathcal R_{1 \sigma}(m_{1,i},m_{2,j}) } \right]^2 \nonumber \\  
&+& \sum_{{\rm bins \, k}}\left[ \frac{R_{\rm EB/LB} (f_{\rm PBH},m_{{\rm ch},k}) \Delta m_{\rm ch} } {  m_{{\rm ch}, k} \times \mathcal R_{1 \sigma}(m_{{\rm ch}, k})} \right]^2
\eea
that combines the ($1\sigma$) rate limits in all the mass bins of the present search and the limits in the chirp mass bins of the previous O2 search~\cite{LIGOScientific:2019kan}.  $R_{\rm EB/LB} (f_{\rm PBH},m_{{\rm ch},k})$ denotes the integrated merger rate per unit logarithmic chirp mass. A PBH population-specific analysis, using injections drawn from, or reweighted to, the predicted PBH binary distribution of each model could have been a more accurate analysis for the integrated merger rate.  However, this would have been much more computationally expensive.  Given the relatively large theoretical uncertainties on the rate limits, this was beyond the scope of the present paper.   For $\gamma = 0.8$, we find that this search greatly improves (by about one order of magnitude) the limits on $f_{\rm PBH}$, compared to previous sub-solar searches.  But as expected, the improvement is marginal if the peak of the mass function lies well below $2 M_\odot$, as in the case $\gamma =0.2$.  For our benchmark modelling of the rates, the combined 90\% CL limits on $f_{\rm PBH}$ lie between $10$ and $100$. The limits beyond the physical range, $f_{\rm PBH} > 1$, do not correspond to physically allowed abundances of PBHs. O2 searches for compact binaries with at least one SSM component lacked the sensitivity needed to constrain $f^{}_{\rm PBH}$ below unity or to place physically meaningful limits on the PBH abundance within various assumptions about PBH model and merger prescriptions. As a consequence, probing $f_{\rm PBH} <1$ will be possible with an improved detector sensitivity by a factor two to four\footnote{Subsequent LVK analyses of SSM on O3 and O4a data~\cite{LIGOScientific:2022hai,LIGOScientific:2026wxz}, which incorporated the formalism presented here, have shown that $f^{}_{\rm PBH}$ can be constrained for EB to values below unity. In contrast, for the LB scenario, $f^{}_{\rm PBH}$ remains unconstrained throughout most of the parameter space. The tighter constraints are largely attributed to improved detector sensitivity~\cite{aLIGO:2020wna,Virgo:2022ysc,Capote:2024rmo,VIRGO:2026okv} compared to O2.}. Other spectral shapes for the primordial fluctuations  (lognormal, broken power-law...) are possible but typically lead to a suppression of low mass ratio binaries compared to a scale-invariant spectrum.  Therefore, the derived limits are relatively conservative.

Finally, we have reanalized the O2 limits on $f_{\rm PBH}$  for monochromatic mass models in~\cite{LIGOScientific:2019kan} with our merging rate prescriptions.  Due to the EB rate suppression, these limits are much less stringent and we find that O2 data do not yet exclude $f_{\rm PBH} = 1$, both for EBs and LBs, as shown in Figure 3 of the Supplementary Material.

\section{Conclusion}
\label{sec:conclusion}
Broad-mass PBH models with a peak at the solar-mass scale from the QCD transition have been proposed to explain the DM and at least some GW events. Motivated by these models, an extended search of mergers of SSM BHs between $0.19$ and $1 M_\odot$ in binaries with a primary component between $1.95$ and 11 $M_\odot$ has been performed on the data from the second observing run (O2) of Advanced LIGO.  The search has revealed four candidates (two being single detector triggers) with a total SNR $>8$ and a FAR $<2 \, {\rm yr}^{-1}$, but whose statistical significance is not high enough to claim for a detection. Therefore, we assume a null results of the search for estimating the merger rate limits.

Using simulated GW signal injections, we estimate the sensitive volume-time of the search, $\langle {\rm VT} \rangle$ at 15 fixed component-mass pairs and derive PBH-model-independent 90\% confidence upper limits on the merger rate, $\mathcal{R}_{90\%}$. These limits extend previous O2 constraints to binaries with heavier primary components and low mass ratios. Their conversion into constraints on $f_{\rm PBH}$ requires assumptions about the PBH mass function, binary-formation channel, early-binary suppression, and late-binary clustering.

When observational limits on merger rates are compared to the theoretical predictions for PBH merging rates for the broad, QCD-motivated mass function and merger rate prescriptions considered here, we find that our limits on $f_{\rm PBH}$ remain above unity, $f_{\rm PBH} > 1$ for both early and late binaries (respectively PBH binaries formed before matter-radiation equality and in PBH clusters). A much weaker constraint is obtained for $f_{\rm PBH}$, when another broad PBH model from De Luca {\it et al.}~\cite{DeLucae:2020agl} is considered, which predicts a much lower merger rate of binaries in the subsolar mass range. The O2 observations therefore do not constrain the PBH abundance within its physical range. This seems to contradict some conclusions of~\cite{LIGOScientific:2019kan}
where more stringent limits on $f_{\rm PBH}$ were obtained. The main difference comes from the inclusion of the rate suppression of early binaries seen in N-body simulations~\cite{Raidal:2018bbj}. The rate suppression included here substantially reduces the predicted rate for binaries formed in the early-binary channel.  

Nevertheless, we emphasize that rate predictions are within the range of the O3 or future O4 and O5 observing runs.  SSM object searches in these data sets should, therefore, ideally include binaries with low mass ratios. Subsequent, LIGO-Virgo-KAGRA analyses of O3~\cite{LIGOScientific:2022hai} and the first part of O4~\cite{LIGOScientific:2026wxz} have included a comparable asymmetric mass ranges in their search space and obtained progressively stronger constraints on $f^{}_{\rm PBH}$ from the early-binary PBH scenario. These demonstrate the gain from increased detector sensitivity and support the continued inclusion of low-mass-ratio systems in searches for SSM compact objects. 

Ultimately, detecting a SSM black hole is the best way to distinguish primordial and stellar BHs. A statistically significant detection of an SSM black hole would provide strong evidence for a nonstandard formation channel of BHs and would have important implications for compact-object astrophysics, cosmology, and fundamental physics. The present search establishes model-independent observational limits for SSM black holes in this asymmetric mass region and motivates its continued exploration with more sensitive data.

\acknowledgments{K.S.P. acknowledges support from STFC grant ST/V005677/1. S.C. and K. S. P. are supported by the research program of the Netherlands Organisation  for  Scientific  Research  (NWO). G.B. is supported by a FRIA grant from the Belgian fund for research FNRS-F.R.S. S.C. aknowledges support from the FNRS-F.R.S and the Franqui foundation.
This research has made use of data, software and/or web tools obtained from the Gravitational Wave Open Science Center (https://www.gw-openscience.org/ ), a service of LIGO Laboratory, the LIGO Scientific Collaboration and the Virgo Collaboration. LIGO Laboratory and Advanced LIGO are funded by the United States National Science Foundation (NSF) as well as the Science and Technology Facilities Council (STFC) of the United Kingdom, the Max-Planck-Society (MPS), and the State of Niedersachsen/Germany for support of the construction of Advanced LIGO and construction and operation of the GEO600 detector. Additional support for Advanced LIGO was provided by the Australian Research Council. Virgo is funded, through the European Gravitational Observatory (EGO), by the French Centre National de Recherche Scientifique (CNRS), the Italian Istituto Nazionale della Fisica Nucleare (INFN) and the Dutch Nikhef, with contributions by institutions from Belgium, Germany, Greece, Hungary, Ireland, Japan, Monaco, Poland, Portugal, Spain. The authors are grateful for computational resources provided by the LIGO Laboratory  and supported by National Science Foundation Grants PHY-0757058 and PHY-0823459.}




\def\ga{\mathrel{\raise.3ex\hbox{$>$\kern-.75em\lower1ex\hbox{$\sim$}}}}
\def\la{\mathrel{\raise.3ex\hbox{$<$\kern-.75em\lower1ex\hbox{$\sim$}}}}

\def\be{\begin{equation}}
\def\ee{\end{equation}}
\def\bea{\begin{eqnarray}}
\def\eea{\end{eqnarray}}

\def\betap{\tilde\beta}
\def\del{\delta_{\rm PBH}^{\rm local}}
\def\Msun{M_\odot}

\title

\begin{center}
{\large\bf
Supplemental materials: Search for sub-solar primordial black holes in low mass ratio binaries with LIGO-Virgo O2 data and implications for the primordial black hole dark matter fraction} 
\end{center}


\section{EB merger rate suppression  factor $f_{\rm sup}$}

For the suppression factor $f_{\rm sup}[f_{\rm PBH},f(m_1),f(m_2)]$ in the EB merging rates of Eq.~\ref{eq:cosmomerg} we have used the analytical prescriptions from Ref.~\cite{Hutsi:2020sol}, adapted to our broad PBH mass distributions, with two contributions
\be
f_{\rm sup} = S_1 \times S_2~.
\ee
The first factor is given by
\be
S_1 = 1.42 \left[ \frac{(\langle m_{\rm PBH}^2 \rangle/\langle m_{\rm PBH} \rangle^2)}{\bar N + C} + \frac{\sigma_{\rm M}^2}{f_{\rm PBH}^2}\right]^{-21/74} {\rm e}^{-\bar N} . \label{eq:S1}
\ee
It takes into account the binary disruption by both matter fluctuations with a (rescaled) variance $\sigma_{\rm M}^2 \simeq 0.005$, and by the number of nearby PBHs $\bar N$  at a  distance smaller than the maximal comoving distance for such  nearby PBH to fall onto the binary before matter-radiation equality.  This number, for any mass function, has been estimated as
\be
\bar N = \frac{m_1+m_2}{\langle m_{\rm PBH} \rangle} \frac{f_{\rm PBH}}{f_{\rm PBH}+\sigma_M}~.  \label{eq:Nbar}
\ee
In Eqs.~(\ref{eq:S1}) and~(\ref{eq:Nbar}), $\langle m_{\rm PBH} \rangle$ is the mean PBH mass  and $\langle m_{\rm PBH}^2 \rangle$ is the corresponding variance.
The function $C$ in Eq.~\ref{eq:S1} encodes the transition between small and large $\bar N$ limits.  A good approximation (with an estimated accuracy of 7\%~\cite{Hutsi:2020sol} compared to the N-body simulations of~\cite{Raidal:2018bbj} for a monochromatic or a lognormal mass function) is given by
\bea
C &\simeq& \frac{f_{\rm PBH}^2 \langle m_{\rm PBH}^2 \rangle} {\sigma_{\rm M}^2 \langle m_{\rm PBH} \rangle^2}  \nonumber \\
& \times & \left\{ \left[ \frac{\Gamma(29/37)}{\sqrt{\pi} } U\left( \frac{21}{74},\frac{1}{2} , \frac{5 f_{\rm PBH}^2}{6 \sigma_{\rm M}^2}\right) \right]^{-74/21}  -1 \right\}^{-1}
\eea
where $\Gamma$ is the Euler function and $U$ is the confluent hypergeometric function.
For an extended mass function spanning several decades of masses, this approach generically gives $\bar N \gg 1$, which would lead to a huge rate suppression for EBs and, as a consequence, LB mergers would be dominant.  However it is unrealistic because it is unlikely that PBHs with a much smaller mass than the one of the binary components can ionize this binary.  Instead, given that we have a sharp peak in the mass function, we  use the approximation $\bar N \simeq 2$ and $\langle m_{\rm PBH}^2 \rangle/\langle m_{\rm PBH} \rangle^2 \simeq 1 $ that corresponds to the monochromatic limit.  
For $f_{\rm PBH} = 1$, one gets $S_1 \approx 0.2$. 
Given the uncertainties on the exact value of $\bar N$, we estimate the possible error on $S_1$ as the following.  As a lower bound, we consider a value of $S_1$ ten times lower than for $\bar N = 2$.  As an upper bound, we consider the limit $\bar N \rightarrow 0$, which gives
\be
S_1^{\rm max} = \left( \frac{5 f_{\rm PBH}^2}{6 \sigma_{\rm M}^2}\right)^{\frac{21}{74}} U\left( \frac{21}{74},\frac{1}{2} , \frac{5 f_{\rm PBH}^2}{6 \sigma_{\rm M}^2}\right)~.
\ee
For $f_{\rm PBH} > 0.1$, one gets $S_1^{\rm max} \simeq 1$ and there is no suppression in the EB merger rate.  These lower and upper bounds on $S_1$ have been used to calculate the uncertainties in the $f_{\rm PBH}$ limits reported in Fig.~\ref{fig:fPBHlimit} for our different mass models.

The second factor $S_2(f_{\rm PBH})$ comes from the binary disruptions in early-forming clusters and can be approximated today by
\be
S_2 \approx \min \left(1,9.6 \times 10^{-3} f_{\rm PBH}^{-0.65} {\rm e}^{0.03 \ln^2 f_{\rm PBH}} \right).
\ee
This factor is inherently conservative, as $S_2$ is designed to  provide a maximal suppression of mergers of initial binaries due to  disruptions in clusters and therefore yields conservative (least stringent) constraints on $f^{}_{\rm PBH}$. Since we are interested by subsolar BHs at typical distances that are still relatively small compared to cosmological scales, one can safely neglect the redshift dependence in $S_2$ from~\cite{Hutsi:2020sol}. For $f_{\rm PBH} = 1$, one gets $S_2 \simeq 0.01$, such that one gets $f_{\rm sup} \simeq 0.002$.  Such a value is also motivated observationally, since it can reproduce the merging rates of GW190521, GW190814 and GW190425~\cite{Clesse:2020ghq}.

Finally, we point out that if $f_{\rm PBH} \sim 1$, the merger rates from perturbed binaries could exceed the rate of non-perturbed binaries~\cite{Vaskonen:2019jpv}, however there is no simple prescription that could be used for extended mass functions and these rates also have large uncertainties.  Therefore we did not consider perturbed binaries in our work.  It can be also pointed out that subtle general relativistic effects may also highly suppress the rate of EBs~\cite{Boehm:2020jwd} if the metric around each PBH is similar to the Thakurta metric.  However this is not the case for a perturbed Mc Vittie metric, as shown in~\cite{DeLuca:2020jug,Hutsi:2021nvs}.

\begin{figure}[t]
\centering
\includegraphics[width = 0.49 \textwidth]{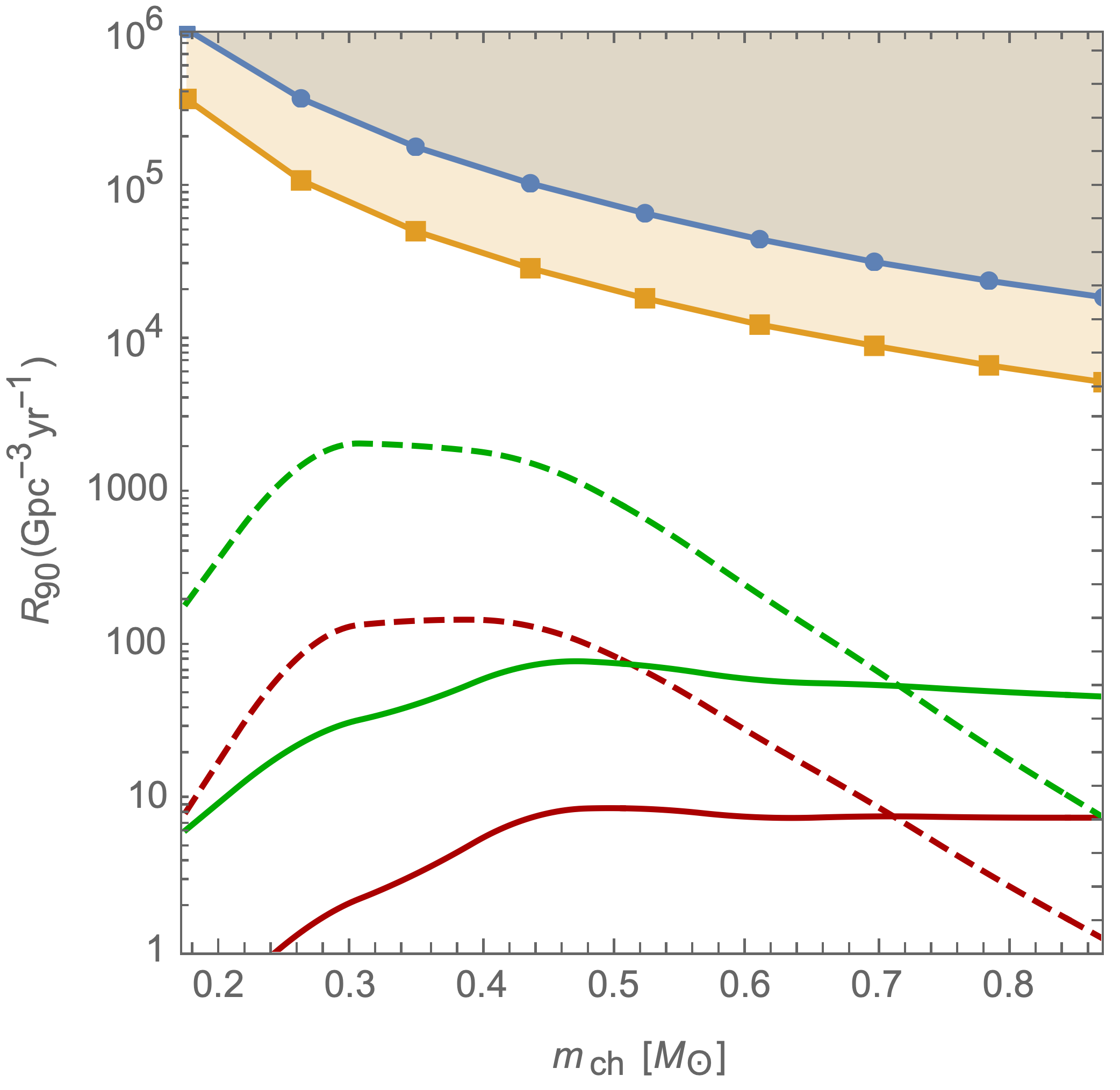}
\caption{\label{fig:limit} The previous O1 (blue line) and O2 (orange line) 90\% C.L. limits on the merger rate as a function of the binary chirp mass~\cite{abbott2018effects}.  Rate predictions for EBs (green) and LBs (red) for the mass model 1 and $f_{\rm PBH} = 1$, with $\gamma = 0.8$ (solid lines) and $0.2$ (dashed lines).} 
\end{figure}

\begin{figure}[t]
\centering
\includegraphics[width=0.49\textwidth]{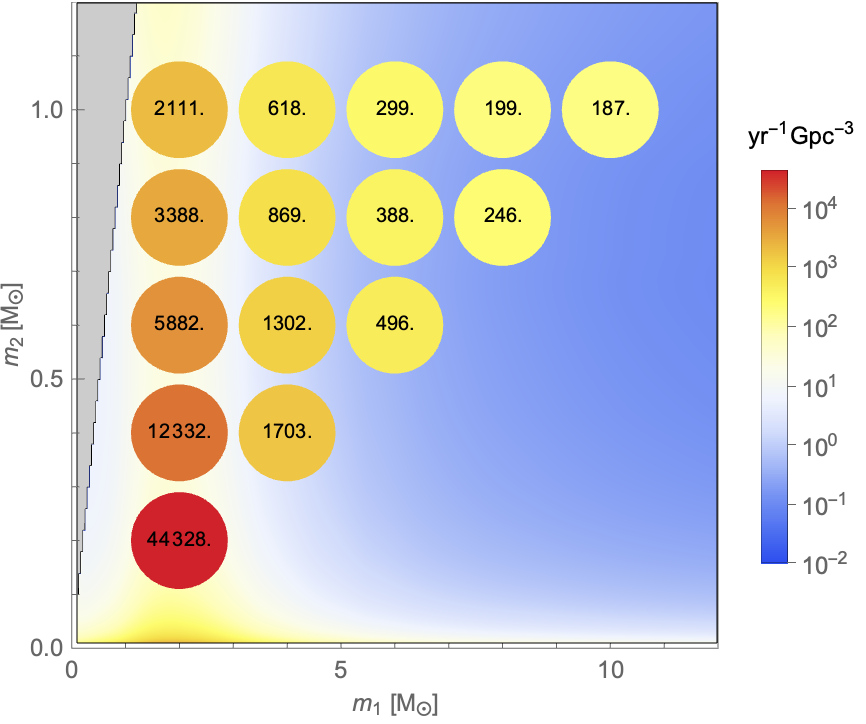}
\caption{\label{fig:2Dlimit_2}
Merger rate predictions for LBs and the mass model 1 with $\gamma = 0.8$ and $f_{\rm PBH} = 1$ in mass bin sizes $\Delta m_1 = 2 M_\odot$ and $\Delta m_2 = 0.2 M_\odot$.  The colored disks represent the $90\%$ rate limits from this extended search in O2 data, as in Fig.~\ref{fig:2Dlimit}.  }
\end{figure}

With these assumptions, the merging rates of EBs and LBs for $f_{\rm PBH} = 1$, $\gamma = 0.8$ or $\gamma = 0.2$ are shown in Figs.~\ref{fig:2Dlimit},~\ref{fig:limit} and~\ref{fig:2Dlimit_2} for the mass model 1.  These rates are compared to the rate limits from the extended search in O2 data.   

\section{Comparison with previous limits}

The rate predictions have also been compared to previous limits restricted to $m_1 < 2M_\odot$~\cite{Abbott:2018oah,LIGOScientific:2019kan}, for both broad-mass and monochromatic models.  On Fig.~\ref{fig:limit} we have represented the rates obtained for the mass model 1 for EBs and LBs, as a function of the binary chirm pass, assuming bins with $\Delta m_{\rm ch} = 0.15 M_\odot$.

\begin{figure}[t]
\centering
\includegraphics[width=0.49\textwidth]{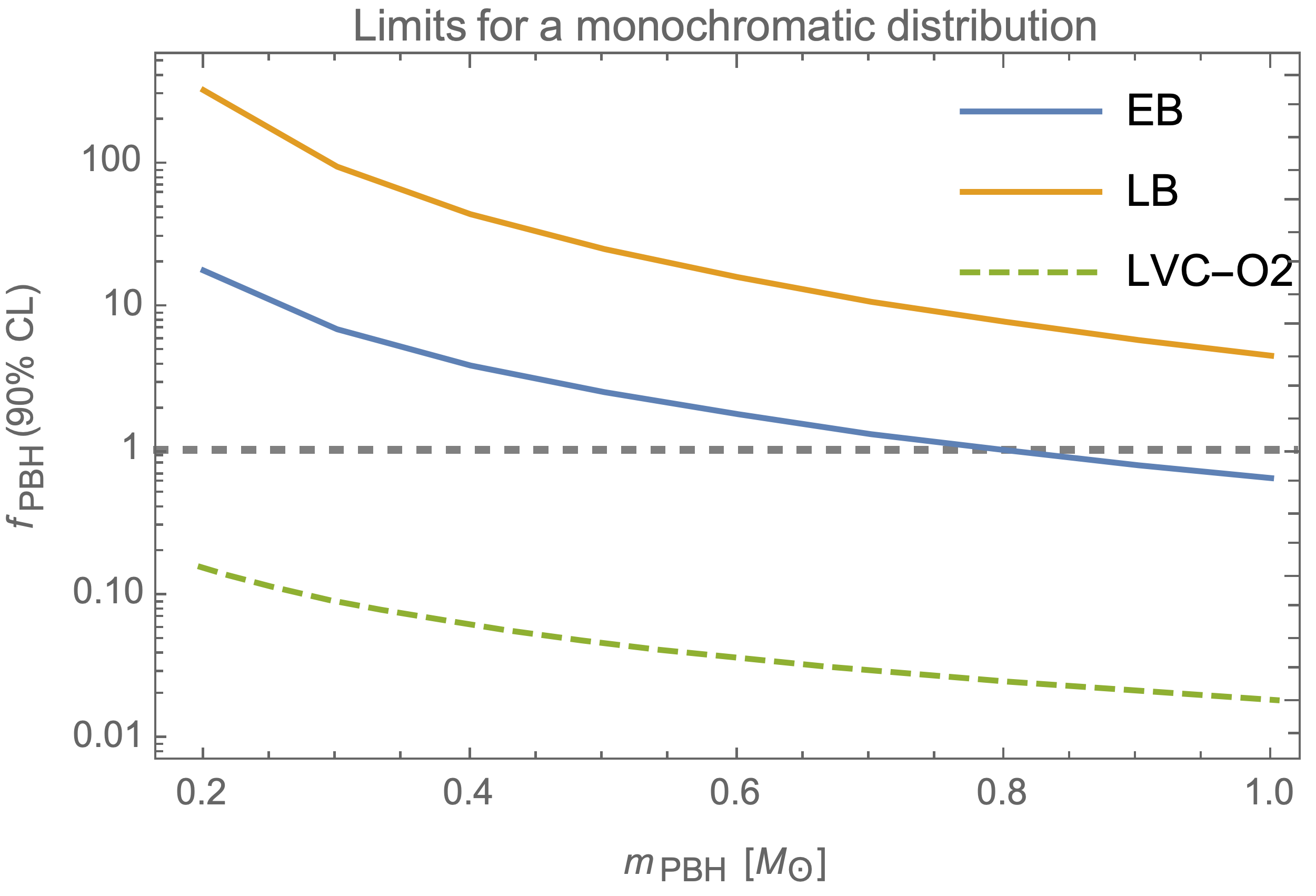}
\caption{Subsolar limits on $f_{\rm PBH}$ for a monochromatic mass model, for late binaries (LB), early binaries (EB) in our benchmark rate models and as estimated for EBs in the previous O2 search~\cite{LIGOScientific:2019kan} without including EB rate suppression effects. } \label{fig:fPBHmono} 
\end{figure}

The previous O2 limits on $f_{\rm PBH}$ for monochromatic mass models can also be re-analysed with our assumptions for the merging rates of EBs and LBs, using the same method as~\cite{LIGOScientific:2019kan}.   These limits are shown in Fig.~\ref{fig:fPBHmono} and for EBs they are two orders of magnitude less stringent than previously found, an effect  due to the rate suppression factor $f_{\rm sup}$.  We find a limit of $f_{\rm PBH}\lesssim 1$ only for EBs and masses between $0.8 M_\odot$ and $1 M_\odot$ but without firmly excluding $f_{\rm PBH}= 1$ given the rate uncertainties.  

\section{Time-frequency analysis}

We perform time-frequency analysis of data around the trigger-times of the four candidates events reported in Table~\ref{tab:candidate} for quantitative inspection about presence of GW signal. The scalograms or q-scans are obtained from  the strain channel data stretches  of the two LIGO detectors centered at trigger times of  the four candidates.  No q-scan provides clear visual indications of chirpy signatures or the presence of GW signal in the time-frequency representation of data.  A similar conclusion is obtained for a search of long-duration bursts 
 80 seconds around the triggers using the strain channel \texttt{DCH-CLEAN\_STRAIN\_C02}, after spectral whitening each 4 seconds with 2 seconds of overlap.  

\begin{figure*}[t]
\centering
\includegraphics[width=0.48\textwidth]{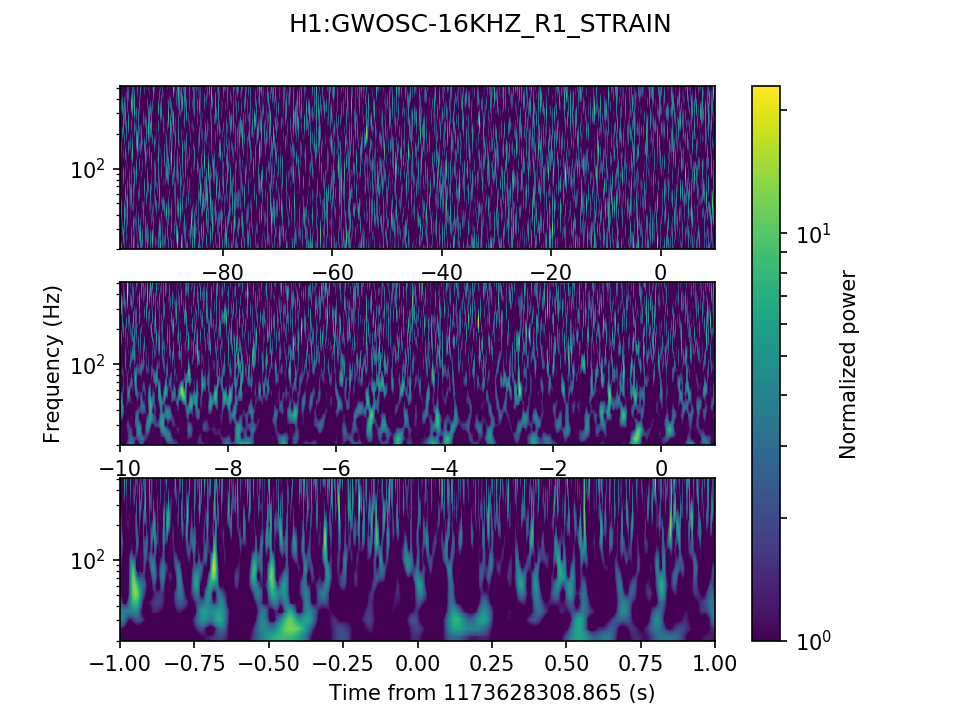}
\includegraphics[width=0.48\textwidth]{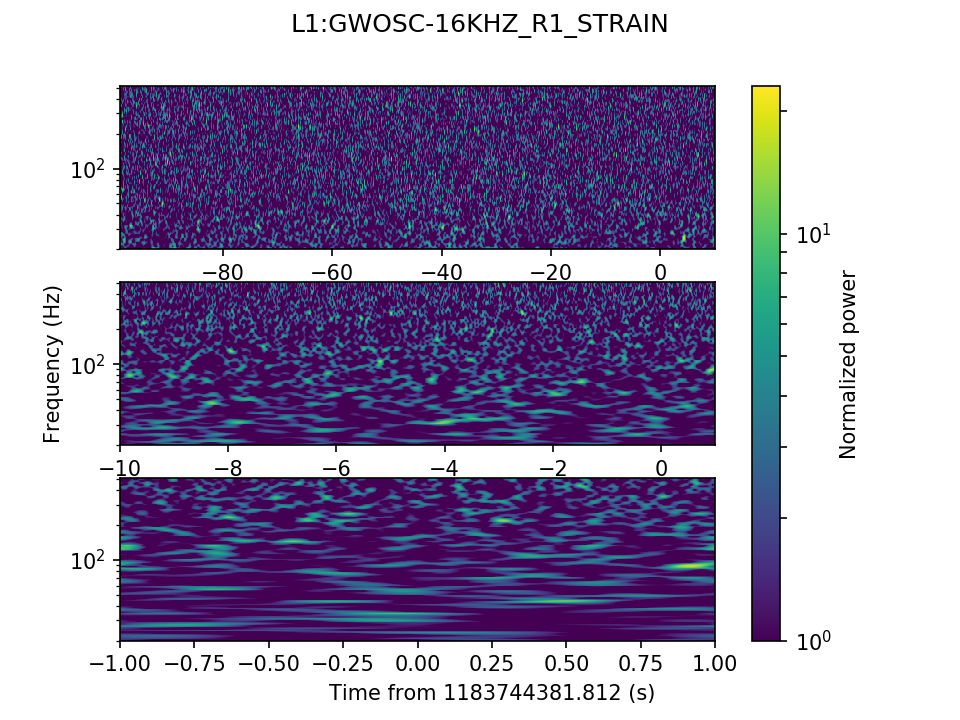}\\
\includegraphics[width=0.48\textwidth]{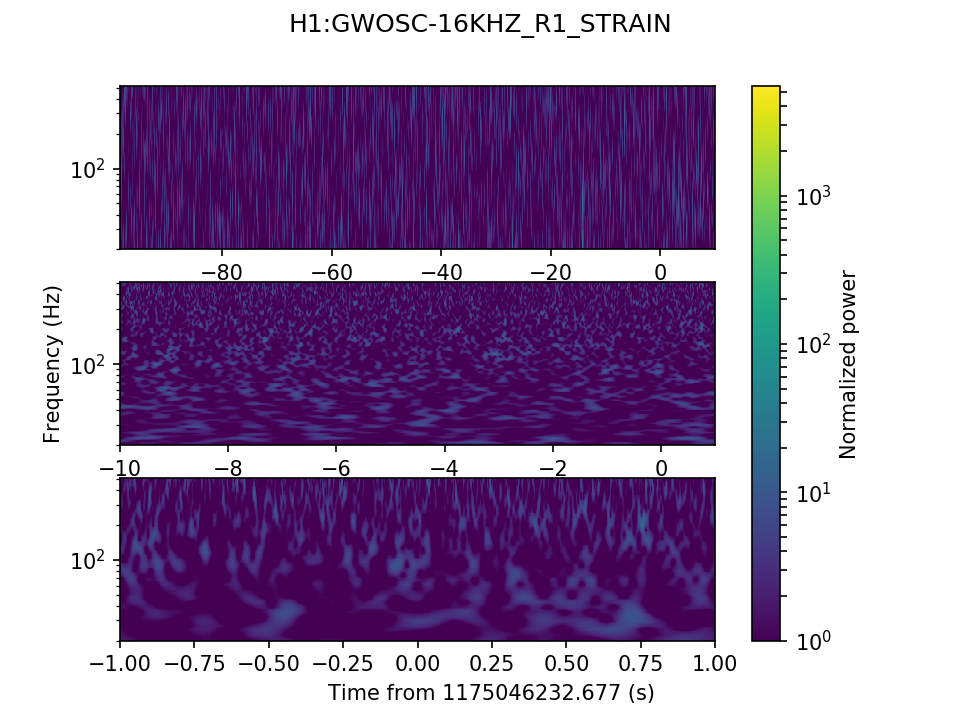}
\includegraphics[width=0.48\textwidth]{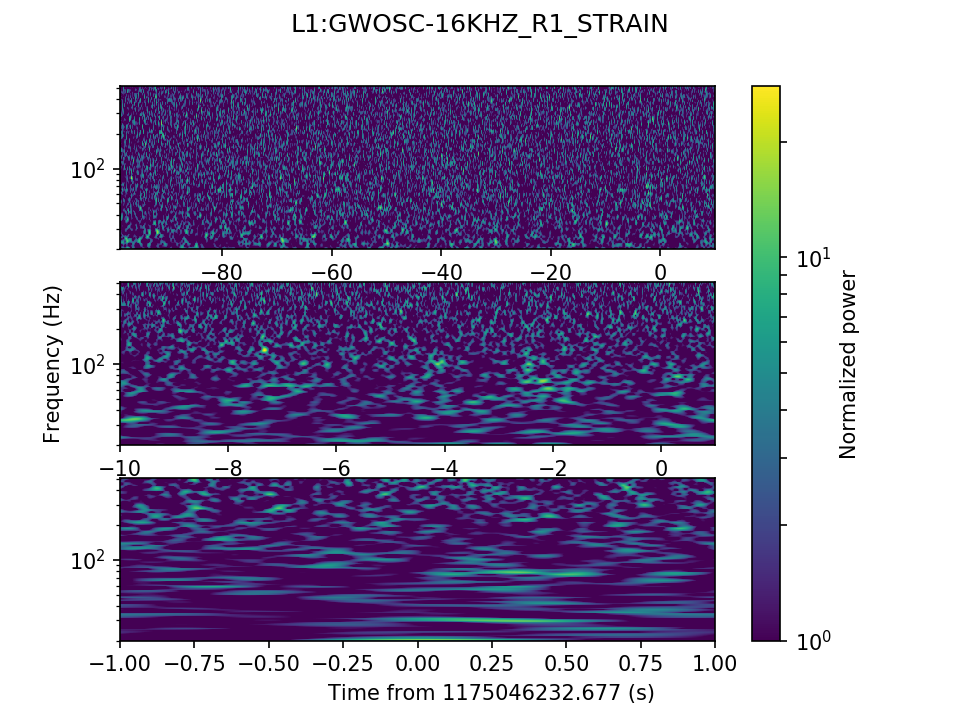}
\\
\includegraphics[width=0.48\textwidth]{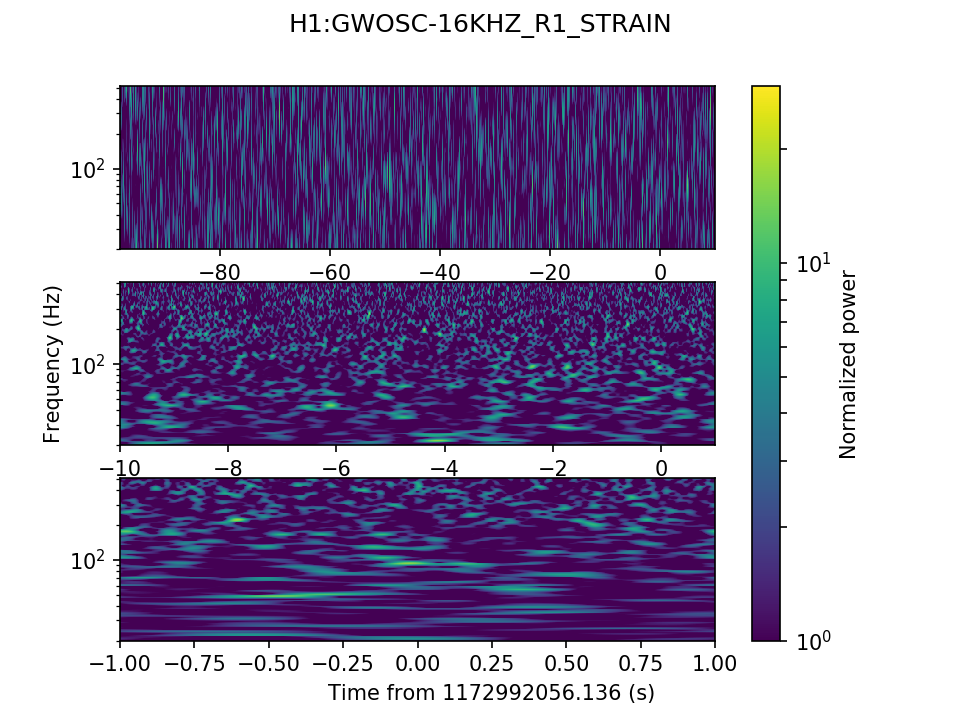} 
\includegraphics[width=0.48\textwidth]{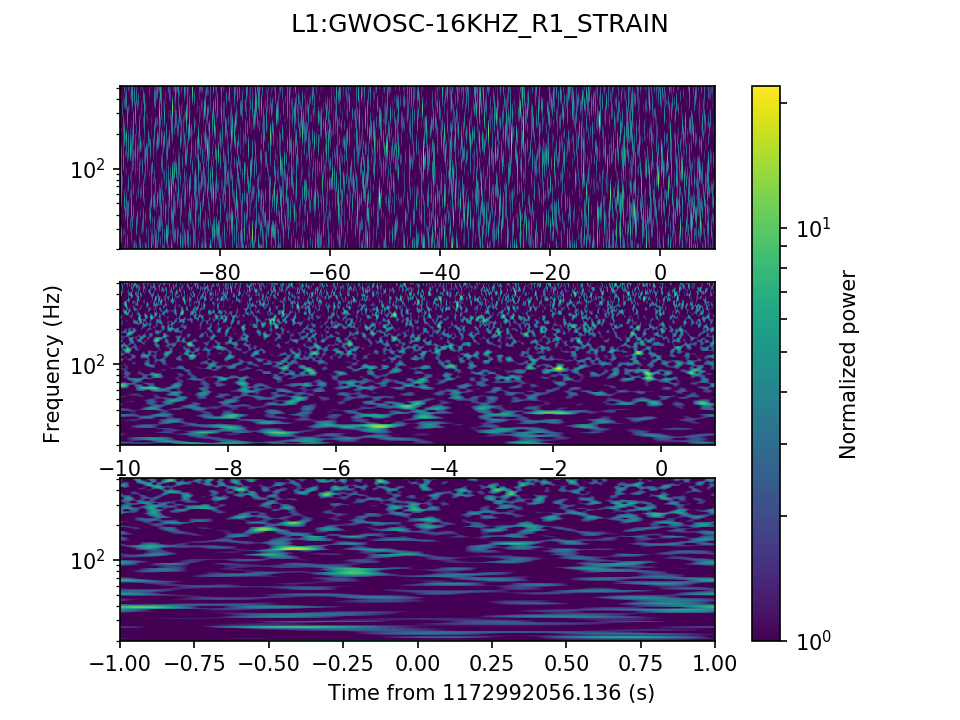}
\caption{\label{fig:omegascans}
Q-scans of strain-channel data centered at  trigger-times of the four candidates of  Table~\ref{tab:candidate}: Candidate 1 in H1 (top left), Candidate 2 in L1 (top right), Candidate 3 (middle panels), Candidate 4 (bottom panels).   For each trigger, three q-scans with different time intervals are shown. The colored scale represents the normalized power. The plots are generated using PyCBC~\cite{alex_nitz_2021_4556907} software. Here, Candidate 1 and Candidate 2 are single detector triggers and observed by LIGO Hanford (H1) and LIGO Livingston (L1), respectively.  Data channel name is displayed on top of each plot.  }
\end{figure*}



\bibliographystyle{apsrev4-1}
\bibliography{PBH_arxiv.bib}

\begin{thebibliography}{116}%
\makeatletter
\providecommand \@ifxundefined [1]{%
 \@ifx{#1\undefined}
}%
\providecommand \@ifnum [1]{%
 \ifnum #1\expandafter \@firstoftwo
 \else \expandafter \@secondoftwo
 \fi
}%
\providecommand \@ifx [1]{%
 \ifx #1\expandafter \@firstoftwo
 \else \expandafter \@secondoftwo
 \fi
}%
\providecommand \natexlab [1]{#1}%
\providecommand \enquote  [1]{``#1''}%
\providecommand \bibnamefont  [1]{#1}%
\providecommand \bibfnamefont [1]{#1}%
\providecommand \citenamefont [1]{#1}%
\providecommand \href@noop [0]{\@secondoftwo}%
\providecommand \href [0]{\begingroup \@sanitize@url \@href}%
\providecommand \@href[1]{\@@startlink{#1}\@@href}%
\providecommand \@@href[1]{\endgroup#1\@@endlink}%
\providecommand \@sanitize@url [0]{\catcode `\\12\catcode `\$12\catcode `\&12\catcode `\#12\catcode `\^12\catcode `\_12\catcode `\%12\relax}%
\providecommand \@@startlink[1]{}%
\providecommand \@@endlink[0]{}%
\providecommand \url  [0]{\begingroup\@sanitize@url \@url }%
\providecommand \@url [1]{\endgroup\@href {#1}{\urlprefix }}%
\providecommand \urlprefix  [0]{URL }%
\providecommand \Eprint [0]{\href }%
\providecommand \doibase [0]{http://dx.doi.org/}%
\providecommand \selectlanguage [0]{\@gobble}%
\providecommand \bibinfo  [0]{\@secondoftwo}%
\providecommand \bibfield  [0]{\@secondoftwo}%
\providecommand \translation [1]{[#1]}%
\providecommand \BibitemOpen [0]{}%
\providecommand \bibitemStop [0]{}%
\providecommand \bibitemNoStop [0]{.\EOS\space}%
\providecommand \EOS [0]{\spacefactor3000\relax}%
\providecommand \BibitemShut  [1]{\csname bibitem#1\endcsname}%
\let\auto@bib@innerbib\@empty
\bibitem [{\citenamefont {Abbott}\ \emph {et~al.}(2016{\natexlab{a}})\citenamefont {Abbott} \emph {et~al.}}]{Abbott:2016blz}%
  \BibitemOpen
  \bibfield  {author} {\bibinfo {author} {\bibfnamefont {B.~P.}\ \bibnamefont {Abbott}} \emph {et~al.} (\bibinfo {collaboration} {LIGO Scientific, Virgo}),\ }\href {\doibase 10.1103/PhysRevLett.116.061102} {\bibfield  {journal} {\bibinfo  {journal} {Phys. Rev. Lett.}\ }\textbf {\bibinfo {volume} {116}},\ \bibinfo {pages} {061102} (\bibinfo {year} {2016}{\natexlab{a}})},\ \Eprint {http://arxiv.org/abs/1602.03837} {arXiv:1602.03837 [gr-qc]} \BibitemShut {NoStop}%
\bibitem [{\citenamefont {Abbott}\ \emph {et~al.}(2020{\natexlab{a}})\citenamefont {Abbott} \emph {et~al.}}]{Abbott:2020tfl}%
  \BibitemOpen
  \bibfield  {author} {\bibinfo {author} {\bibfnamefont {R.}~\bibnamefont {Abbott}} \emph {et~al.} (\bibinfo {collaboration} {LIGO Scientific, Virgo}),\ }\href {\doibase 10.1103/PhysRevLett.125.101102} {\bibfield  {journal} {\bibinfo  {journal} {Phys. Rev. Lett.}\ }\textbf {\bibinfo {volume} {125}},\ \bibinfo {pages} {101102} (\bibinfo {year} {2020}{\natexlab{a}})},\ \Eprint {http://arxiv.org/abs/2009.01075} {arXiv:2009.01075 [gr-qc]} \BibitemShut {NoStop}%
\bibitem [{\citenamefont {Abbott}\ \emph {et~al.}(2020{\natexlab{b}})\citenamefont {Abbott} \emph {et~al.}}]{Abbott:2020khf}%
  \BibitemOpen
  \bibfield  {author} {\bibinfo {author} {\bibfnamefont {R.}~\bibnamefont {Abbott}} \emph {et~al.} (\bibinfo {collaboration} {LIGO Scientific, Virgo}),\ }\href {\doibase 10.3847/2041-8213/ab960f} {\bibfield  {journal} {\bibinfo  {journal} {Astrophys. J. Lett.}\ }\textbf {\bibinfo {volume} {896}},\ \bibinfo {pages} {L44} (\bibinfo {year} {2020}{\natexlab{b}})},\ \Eprint {http://arxiv.org/abs/2006.12611} {arXiv:2006.12611 [astro-ph.HE]} \BibitemShut {NoStop}%
\bibitem [{\citenamefont {Abbott}\ \emph {et~al.}(2020{\natexlab{c}})\citenamefont {Abbott} \emph {et~al.}}]{LIGOScientific:2020stg}%
  \BibitemOpen
  \bibfield  {author} {\bibinfo {author} {\bibfnamefont {R.}~\bibnamefont {Abbott}} \emph {et~al.} (\bibinfo {collaboration} {LIGO Scientific, Virgo}),\ }\href {\doibase 10.1103/PhysRevD.102.043015} {\bibfield  {journal} {\bibinfo  {journal} {Phys. Rev. D}\ }\textbf {\bibinfo {volume} {102}},\ \bibinfo {pages} {043015} (\bibinfo {year} {2020}{\natexlab{c}})},\ \Eprint {http://arxiv.org/abs/2004.08342} {arXiv:2004.08342 [astro-ph.HE]} \BibitemShut {NoStop}%
\bibitem [{\citenamefont {Abbott}\ \emph {et~al.}(2017{\natexlab{a}})\citenamefont {Abbott} \emph {et~al.}}]{Abbott:2017oio}%
  \BibitemOpen
  \bibfield  {author} {\bibinfo {author} {\bibfnamefont {B.~P.}\ \bibnamefont {Abbott}} \emph {et~al.} (\bibinfo {collaboration} {LIGO Scientific, Virgo}),\ }\href {\doibase 10.1103/PhysRevLett.119.141101} {\bibfield  {journal} {\bibinfo  {journal} {Phys. Rev. Lett.}\ }\textbf {\bibinfo {volume} {119}},\ \bibinfo {pages} {141101} (\bibinfo {year} {2017}{\natexlab{a}})},\ \Eprint {http://arxiv.org/abs/1709.09660} {arXiv:1709.09660 [gr-qc]} \BibitemShut {NoStop}%
\bibitem [{\citenamefont {Abbott}\ \emph {et~al.}(2016{\natexlab{b}})\citenamefont {Abbott} \emph {et~al.}}]{Abbott:2016nmj}%
  \BibitemOpen
  \bibfield  {author} {\bibinfo {author} {\bibfnamefont {B.~P.}\ \bibnamefont {Abbott}} \emph {et~al.} (\bibinfo {collaboration} {LIGO Scientific, Virgo}),\ }\href {\doibase 10.1103/PhysRevLett.116.241103} {\bibfield  {journal} {\bibinfo  {journal} {Phys. Rev. Lett.}\ }\textbf {\bibinfo {volume} {116}},\ \bibinfo {pages} {241103} (\bibinfo {year} {2016}{\natexlab{b}})},\ \Eprint {http://arxiv.org/abs/1606.04855} {arXiv:1606.04855 [gr-qc]} \BibitemShut {NoStop}%
\bibitem [{\citenamefont {Abbott}\ \emph {et~al.}(2017{\natexlab{b}})\citenamefont {Abbott} \emph {et~al.}}]{Abbott:2017gyy}%
  \BibitemOpen
  \bibfield  {author} {\bibinfo {author} {\bibfnamefont {B.~. P.~.}\ \bibnamefont {Abbott}} \emph {et~al.} (\bibinfo {collaboration} {LIGO Scientific, Virgo}),\ }\href {\doibase 10.3847/2041-8213/aa9f0c} {\bibfield  {journal} {\bibinfo  {journal} {Astrophys. J.}\ }\textbf {\bibinfo {volume} {851}},\ \bibinfo {pages} {L35} (\bibinfo {year} {2017}{\natexlab{b}})},\ \Eprint {http://arxiv.org/abs/1711.05578} {arXiv:1711.05578 [astro-ph.HE]} \BibitemShut {NoStop}%
\bibitem [{\citenamefont {Abbott}\ \emph {et~al.}(2017{\natexlab{c}})\citenamefont {Abbott} \emph {et~al.}}]{Abbott:2017vtc}%
  \BibitemOpen
  \bibfield  {author} {\bibinfo {author} {\bibfnamefont {B.~P.}\ \bibnamefont {Abbott}} \emph {et~al.} (\bibinfo {collaboration} {LIGO Scientific, VIRGO}),\ }\href {\doibase 10.1103/PhysRevLett.118.221101} {\bibfield  {journal} {\bibinfo  {journal} {Phys. Rev. Lett.}\ }\textbf {\bibinfo {volume} {118}},\ \bibinfo {pages} {221101} (\bibinfo {year} {2017}{\natexlab{c}})},\ \bibinfo {note} {[Erratum: Phys.Rev.Lett. 121, 129901 (2018)]},\ \Eprint {http://arxiv.org/abs/1706.01812} {arXiv:1706.01812 [gr-qc]} \BibitemShut {NoStop}%
\bibitem [{\citenamefont {Abbott}\ \emph {et~al.}(2019{\natexlab{a}})\citenamefont {Abbott} \emph {et~al.}}]{LIGOScientific:2018mvr}%
  \BibitemOpen
  \bibfield  {author} {\bibinfo {author} {\bibfnamefont {B.~P.}\ \bibnamefont {Abbott}} \emph {et~al.} (\bibinfo {collaboration} {LIGO Scientific, Virgo}),\ }\href {\doibase 10.1103/PhysRevX.9.031040} {\bibfield  {journal} {\bibinfo  {journal} {Phys. Rev. X}\ }\textbf {\bibinfo {volume} {9}},\ \bibinfo {pages} {031040} (\bibinfo {year} {2019}{\natexlab{a}})},\ \Eprint {http://arxiv.org/abs/1811.12907} {arXiv:1811.12907 [astro-ph.HE]} \BibitemShut {NoStop}%
\bibitem [{\citenamefont {Abbott}\ \emph {et~al.}(2020{\natexlab{d}})\citenamefont {Abbott} \emph {et~al.}}]{Abbott:2020niy}%
  \BibitemOpen
  \bibfield  {author} {\bibinfo {author} {\bibfnamefont {R.}~\bibnamefont {Abbott}} \emph {et~al.} (\bibinfo {collaboration} {LIGO Scientific, Virgo}),\ }\href@noop {} {\bibfield  {journal} {\bibinfo  {journal} {arXiv}\ } (\bibinfo {year} {2020}{\natexlab{d}})},\ \Eprint {http://arxiv.org/abs/2010.14527} {arXiv:2010.14527 [gr-qc]} \BibitemShut {NoStop}%
\bibitem [{\citenamefont {Abbott}\ \emph {et~al.}(2016{\natexlab{c}})\citenamefont {Abbott} \emph {et~al.}}]{TheLIGOScientific:2016pea}%
  \BibitemOpen
  \bibfield  {author} {\bibinfo {author} {\bibfnamefont {B.~P.}\ \bibnamefont {Abbott}} \emph {et~al.} (\bibinfo {collaboration} {LIGO Scientific, Virgo}),\ }\href {\doibase 10.1103/PhysRevX.6.041015} {\bibfield  {journal} {\bibinfo  {journal} {Phys. Rev. X}\ }\textbf {\bibinfo {volume} {6}},\ \bibinfo {pages} {041015} (\bibinfo {year} {2016}{\natexlab{c}})},\ \bibinfo {note} {[Erratum: Phys.Rev.X 8, 039903 (2018)]},\ \Eprint {http://arxiv.org/abs/1606.04856} {arXiv:1606.04856 [gr-qc]} \BibitemShut {NoStop}%
\bibitem [{\citenamefont {Abac}\ \emph {et~al.}(2026{\natexlab{a}})\citenamefont {Abac} \emph {et~al.}}]{LIGOScientific:2026wfs}%
  \BibitemOpen
  \bibfield  {author} {\bibinfo {author} {\bibfnamefont {A.~G.}\ \bibnamefont {Abac}} \emph {et~al.} (\bibinfo {collaboration} {LIGO Scientific, VIRGO, KAGRA}),\ }\href@noop {} {\enquote {\bibinfo {title} {{GWTC-5.0: Observations from the Second Part of the Fourth LIGO-Virgo-KAGRA Observing Run and Updates to the Gravitational-Wave Transient Catalog}},}\ } (\bibinfo {year} {2026}{\natexlab{a}}),\ \Eprint {http://arxiv.org/abs/2605.27225} {arXiv:2605.27225 [gr-qc]} \BibitemShut {NoStop}%
\bibitem [{\citenamefont {Abac}\ \emph {et~al.}(2026{\natexlab{b}})\citenamefont {Abac} \emph {et~al.}}]{LIGOScientific:2026ctl}%
  \BibitemOpen
  \bibfield  {author} {\bibinfo {author} {\bibfnamefont {A.~G.}\ \bibnamefont {Abac}} \emph {et~al.} (\bibinfo {collaboration} {LIGO Scientific, VIRGO, KAGRA}),\ }\href@noop {} {\enquote {\bibinfo {title} {{GWTC-5.0: Population Properties of Merging Compact Binaries}},}\ } (\bibinfo {year} {2026}{\natexlab{b}}),\ \Eprint {http://arxiv.org/abs/2605.27226} {arXiv:2605.27226 [astro-ph.HE]} \BibitemShut {NoStop}%
\bibitem [{\citenamefont {Aasi}\ \emph {et~al.}(2015)\citenamefont {Aasi} \emph {et~al.}}]{TheLIGOScientific:2014jea}%
  \BibitemOpen
  \bibfield  {author} {\bibinfo {author} {\bibfnamefont {J.}~\bibnamefont {Aasi}} \emph {et~al.} (\bibinfo {collaboration} {LIGO Scientific}),\ }\href {\doibase 10.1088/0264-9381/32/7/074001} {\bibfield  {journal} {\bibinfo  {journal} {Class. Quant. Grav.}\ }\textbf {\bibinfo {volume} {32}},\ \bibinfo {pages} {074001} (\bibinfo {year} {2015})},\ \Eprint {http://arxiv.org/abs/1411.4547} {arXiv:1411.4547 [gr-qc]} \BibitemShut {NoStop}%
\bibitem [{\citenamefont {Acernese}\ \emph {et~al.}(2015)\citenamefont {Acernese} \emph {et~al.}}]{TheVirgo:2014hva}%
  \BibitemOpen
  \bibfield  {author} {\bibinfo {author} {\bibfnamefont {F.}~\bibnamefont {Acernese}} \emph {et~al.} (\bibinfo {collaboration} {VIRGO}),\ }\href {\doibase 10.1088/0264-9381/32/2/024001} {\bibfield  {journal} {\bibinfo  {journal} {Class. Quant. Grav.}\ }\textbf {\bibinfo {volume} {32}},\ \bibinfo {pages} {024001} (\bibinfo {year} {2015})},\ \Eprint {http://arxiv.org/abs/1408.3978} {arXiv:1408.3978 [gr-qc]} \BibitemShut {NoStop}%
\bibitem [{\citenamefont {Aso}\ \emph {et~al.}(2013)\citenamefont {Aso}, \citenamefont {Michimura}, \citenamefont {Somiya}, \citenamefont {Ando}, \citenamefont {Miyakawa}, \citenamefont {Sekiguchi}, \citenamefont {Tatsumi},\ and\ \citenamefont {Yamamoto}}]{Aso:2013eba}%
  \BibitemOpen
  \bibfield  {author} {\bibinfo {author} {\bibfnamefont {Y.}~\bibnamefont {Aso}}, \bibinfo {author} {\bibfnamefont {Y.}~\bibnamefont {Michimura}}, \bibinfo {author} {\bibfnamefont {K.}~\bibnamefont {Somiya}}, \bibinfo {author} {\bibfnamefont {M.}~\bibnamefont {Ando}}, \bibinfo {author} {\bibfnamefont {O.}~\bibnamefont {Miyakawa}}, \bibinfo {author} {\bibfnamefont {T.}~\bibnamefont {Sekiguchi}}, \bibinfo {author} {\bibfnamefont {D.}~\bibnamefont {Tatsumi}}, \ and\ \bibinfo {author} {\bibfnamefont {H.}~\bibnamefont {Yamamoto}} (\bibinfo {collaboration} {KAGRA}),\ }\href {\doibase 10.1103/PhysRevD.88.043007} {\bibfield  {journal} {\bibinfo  {journal} {Phys. Rev. D}\ }\textbf {\bibinfo {volume} {88}},\ \bibinfo {pages} {043007} (\bibinfo {year} {2013})},\ \Eprint {http://arxiv.org/abs/1306.6747} {arXiv:1306.6747 [gr-qc]} \BibitemShut {NoStop}%
\bibitem [{\citenamefont {Akutsu}\ \emph {et~al.}(2021)\citenamefont {Akutsu} \emph {et~al.}}]{KAGRA:2020tym}%
  \BibitemOpen
  \bibfield  {author} {\bibinfo {author} {\bibfnamefont {T.}~\bibnamefont {Akutsu}} \emph {et~al.} (\bibinfo {collaboration} {KAGRA}),\ }\href {\doibase 10.1093/ptep/ptaa125} {\bibfield  {journal} {\bibinfo  {journal} {PTEP}\ }\textbf {\bibinfo {volume} {2021}},\ \bibinfo {pages} {05A101} (\bibinfo {year} {2021})},\ \Eprint {http://arxiv.org/abs/2005.05574} {arXiv:2005.05574 [physics.ins-det]} \BibitemShut {NoStop}%
\bibitem [{\citenamefont {Somiya}(2012)}]{Somiya:2011np}%
  \BibitemOpen
  \bibfield  {author} {\bibinfo {author} {\bibfnamefont {K.}~\bibnamefont {Somiya}} (\bibinfo {collaboration} {KAGRA}),\ }\href {\doibase 10.1088/0264-9381/29/12/124007} {\bibfield  {journal} {\bibinfo  {journal} {Class. Quant. Grav.}\ }\textbf {\bibinfo {volume} {29}},\ \bibinfo {pages} {124007} (\bibinfo {year} {2012})},\ \Eprint {http://arxiv.org/abs/1111.7185} {arXiv:1111.7185 [gr-qc]} \BibitemShut {NoStop}%
\bibitem [{\citenamefont {Bird}\ \emph {et~al.}(2016)\citenamefont {Bird}, \citenamefont {Cholis}, \citenamefont {Mu\~noz}, \citenamefont {Ali-Ha\"\i{}moud}, \citenamefont {Kamionkowski}, \citenamefont {Kovetz}, \citenamefont {Raccanelli},\ and\ \citenamefont {Riess}}]{Bird:2016dcv}%
  \BibitemOpen
  \bibfield  {author} {\bibinfo {author} {\bibfnamefont {S.}~\bibnamefont {Bird}}, \bibinfo {author} {\bibfnamefont {I.}~\bibnamefont {Cholis}}, \bibinfo {author} {\bibfnamefont {J.~B.}\ \bibnamefont {Mu\~noz}}, \bibinfo {author} {\bibfnamefont {Y.}~\bibnamefont {Ali-Ha\"\i{}moud}}, \bibinfo {author} {\bibfnamefont {M.}~\bibnamefont {Kamionkowski}}, \bibinfo {author} {\bibfnamefont {E.~D.}\ \bibnamefont {Kovetz}}, \bibinfo {author} {\bibfnamefont {A.}~\bibnamefont {Raccanelli}}, \ and\ \bibinfo {author} {\bibfnamefont {A.~G.}\ \bibnamefont {Riess}},\ }\href {\doibase 10.1103/PhysRevLett.116.201301} {\bibfield  {journal} {\bibinfo  {journal} {Phys. Rev. Lett.}\ }\textbf {\bibinfo {volume} {116}},\ \bibinfo {pages} {201301} (\bibinfo {year} {2016})},\ \Eprint {http://arxiv.org/abs/1603.00464} {arXiv:1603.00464 [astro-ph.CO]} \BibitemShut {NoStop}%
\bibitem [{\citenamefont {Clesse}\ and\ \citenamefont {Garc\'\i{}a-Bellido}(2017)}]{Clesse:2016vqa}%
  \BibitemOpen
  \bibfield  {author} {\bibinfo {author} {\bibfnamefont {S.}~\bibnamefont {Clesse}}\ and\ \bibinfo {author} {\bibfnamefont {J.}~\bibnamefont {Garc\'\i{}a-Bellido}},\ }\href {\doibase 10.1016/j.dark.2016.10.002} {\bibfield  {journal} {\bibinfo  {journal} {Phys. Dark Univ.}\ }\textbf {\bibinfo {volume} {15}},\ \bibinfo {pages} {142} (\bibinfo {year} {2017})},\ \Eprint {http://arxiv.org/abs/1603.05234} {arXiv:1603.05234 [astro-ph.CO]} \BibitemShut {NoStop}%
\bibitem [{\citenamefont {Sasaki}\ \emph {et~al.}(2016)\citenamefont {Sasaki}, \citenamefont {Suyama}, \citenamefont {Tanaka},\ and\ \citenamefont {Yokoyama}}]{Sasaki:2016jop}%
  \BibitemOpen
  \bibfield  {author} {\bibinfo {author} {\bibfnamefont {M.}~\bibnamefont {Sasaki}}, \bibinfo {author} {\bibfnamefont {T.}~\bibnamefont {Suyama}}, \bibinfo {author} {\bibfnamefont {T.}~\bibnamefont {Tanaka}}, \ and\ \bibinfo {author} {\bibfnamefont {S.}~\bibnamefont {Yokoyama}},\ }\href {\doibase 10.1103/PhysRevLett.117.061101} {\bibfield  {journal} {\bibinfo  {journal} {Phys. Rev. Lett.}\ }\textbf {\bibinfo {volume} {117}},\ \bibinfo {pages} {061101} (\bibinfo {year} {2016})},\ \bibinfo {note} {[Erratum: Phys.Rev.Lett. 121, 059901 (2018)]},\ \Eprint {http://arxiv.org/abs/1603.08338} {arXiv:1603.08338 [astro-ph.CO]} \BibitemShut {NoStop}%
\bibitem [{\citenamefont {Zel'dovich}\ and\ \citenamefont {Novikov}(1967)}]{Zeldovich:1967lct}%
  \BibitemOpen
  \bibfield  {author} {\bibinfo {author} {\bibfnamefont {Y.~B.}\ \bibnamefont {Zel'dovich}}\ and\ \bibinfo {author} {\bibfnamefont {I.~D.}\ \bibnamefont {Novikov}},\ }\href@noop {} {\bibfield  {journal} {\bibinfo  {journal} {Soviet Astronomy}\ }\textbf {\bibinfo {volume} {10}},\ \bibinfo {pages} {602} (\bibinfo {year} {1967})}\BibitemShut {NoStop}%
\bibitem [{\citenamefont {Hawking}(1971)}]{Hawking:1971ei}%
  \BibitemOpen
  \bibfield  {author} {\bibinfo {author} {\bibfnamefont {S.}~\bibnamefont {Hawking}},\ }\href@noop {} {\bibfield  {journal} {\bibinfo  {journal} {Mon. Not. Roy. Astron. Soc.}\ }\textbf {\bibinfo {volume} {152}},\ \bibinfo {pages} {75} (\bibinfo {year} {1971})}\BibitemShut {NoStop}%
\bibitem [{\citenamefont {Carr}\ and\ \citenamefont {Hawking}(1974)}]{Carr:1974nx}%
  \BibitemOpen
  \bibfield  {author} {\bibinfo {author} {\bibfnamefont {B.~J.}\ \bibnamefont {Carr}}\ and\ \bibinfo {author} {\bibfnamefont {S.~W.}\ \bibnamefont {Hawking}},\ }\href@noop {} {\bibfield  {journal} {\bibinfo  {journal} {Mon. Not. Roy. Astron. Soc.}\ }\textbf {\bibinfo {volume} {168}},\ \bibinfo {pages} {399} (\bibinfo {year} {1974})}\BibitemShut {NoStop}%
\bibitem [{\citenamefont {{Chapline}}(1975)}]{1975Natur.253..251C}%
  \BibitemOpen
  \bibfield  {author} {\bibinfo {author} {\bibfnamefont {G.~F.}\ \bibnamefont {{Chapline}}},\ }\href {\doibase 10.1038/253251a0} {\bibfield  {journal} {\bibinfo  {journal} {Nature}\ }\textbf {\bibinfo {volume} {253}},\ \bibinfo {pages} {251} (\bibinfo {year} {1975})}\BibitemShut {NoStop}%
\bibitem [{\citenamefont {Carr}\ and\ \citenamefont {Lidsey}(1993)}]{Carr:1993aq}%
  \BibitemOpen
  \bibfield  {author} {\bibinfo {author} {\bibfnamefont {B.~J.}\ \bibnamefont {Carr}}\ and\ \bibinfo {author} {\bibfnamefont {J.~E.}\ \bibnamefont {Lidsey}},\ }\href {\doibase 10.1103/PhysRevD.48.543} {\bibfield  {journal} {\bibinfo  {journal} {Phys. Rev. D}\ }\textbf {\bibinfo {volume} {48}},\ \bibinfo {pages} {543} (\bibinfo {year} {1993})}\BibitemShut {NoStop}%
\bibitem [{\citenamefont {Ivanov}\ \emph {et~al.}(1994)\citenamefont {Ivanov}, \citenamefont {Naselsky},\ and\ \citenamefont {Novikov}}]{Ivanov:1994pa}%
  \BibitemOpen
  \bibfield  {author} {\bibinfo {author} {\bibfnamefont {P.}~\bibnamefont {Ivanov}}, \bibinfo {author} {\bibfnamefont {P.}~\bibnamefont {Naselsky}}, \ and\ \bibinfo {author} {\bibfnamefont {I.}~\bibnamefont {Novikov}},\ }\href {\doibase 10.1103/PhysRevD.50.7173} {\bibfield  {journal} {\bibinfo  {journal} {Phys. Rev. D}\ }\textbf {\bibinfo {volume} {50}},\ \bibinfo {pages} {7173} (\bibinfo {year} {1994})}\BibitemShut {NoStop}%
\bibitem [{\citenamefont {Garcia-Bellido}\ \emph {et~al.}(1996)\citenamefont {Garcia-Bellido}, \citenamefont {Linde},\ and\ \citenamefont {Wands}}]{GarciaBellido:1996qt}%
  \BibitemOpen
  \bibfield  {author} {\bibinfo {author} {\bibfnamefont {J.}~\bibnamefont {Garcia-Bellido}}, \bibinfo {author} {\bibfnamefont {A.~D.}\ \bibnamefont {Linde}}, \ and\ \bibinfo {author} {\bibfnamefont {D.}~\bibnamefont {Wands}},\ }\href {\doibase 10.1103/PhysRevD.54.6040} {\bibfield  {journal} {\bibinfo  {journal} {Phys. Rev. D}\ }\textbf {\bibinfo {volume} {54}},\ \bibinfo {pages} {6040} (\bibinfo {year} {1996})},\ \Eprint {http://arxiv.org/abs/astro-ph/9605094} {arXiv:astro-ph/9605094} \BibitemShut {NoStop}%
\bibitem [{\citenamefont {Kim}\ and\ \citenamefont {Lee}(1996)}]{Kim:1996hr}%
  \BibitemOpen
  \bibfield  {author} {\bibinfo {author} {\bibfnamefont {H.~I.}\ \bibnamefont {Kim}}\ and\ \bibinfo {author} {\bibfnamefont {C.~H.}\ \bibnamefont {Lee}},\ }\href {\doibase 10.1103/PhysRevD.54.6001} {\bibfield  {journal} {\bibinfo  {journal} {Phys. Rev. D}\ }\textbf {\bibinfo {volume} {54}},\ \bibinfo {pages} {6001} (\bibinfo {year} {1996})}\BibitemShut {NoStop}%
\bibitem [{\citenamefont {Carr}\ \emph {et~al.}(2016)\citenamefont {Carr}, \citenamefont {Kuhnel},\ and\ \citenamefont {Sandstad}}]{Carr:2016drx}%
  \BibitemOpen
  \bibfield  {author} {\bibinfo {author} {\bibfnamefont {B.}~\bibnamefont {Carr}}, \bibinfo {author} {\bibfnamefont {F.}~\bibnamefont {Kuhnel}}, \ and\ \bibinfo {author} {\bibfnamefont {M.}~\bibnamefont {Sandstad}},\ }\href {\doibase 10.1103/PhysRevD.94.083504} {\bibfield  {journal} {\bibinfo  {journal} {Phys. Rev. D}\ }\textbf {\bibinfo {volume} {94}},\ \bibinfo {pages} {083504} (\bibinfo {year} {2016})},\ \Eprint {http://arxiv.org/abs/1607.06077} {arXiv:1607.06077 [astro-ph.CO]} \BibitemShut {NoStop}%
\bibitem [{\citenamefont {Carr}\ \emph {et~al.}(2010)\citenamefont {Carr}, \citenamefont {Kohri}, \citenamefont {Sendouda},\ and\ \citenamefont {Yokoyama}}]{Carr:2009jm}%
  \BibitemOpen
  \bibfield  {author} {\bibinfo {author} {\bibfnamefont {B.}~\bibnamefont {Carr}}, \bibinfo {author} {\bibfnamefont {K.}~\bibnamefont {Kohri}}, \bibinfo {author} {\bibfnamefont {Y.}~\bibnamefont {Sendouda}}, \ and\ \bibinfo {author} {\bibfnamefont {J.}~\bibnamefont {Yokoyama}},\ }\href {\doibase 10.1103/PhysRevD.81.104019} {\bibfield  {journal} {\bibinfo  {journal} {Phys. Rev. D}\ }\textbf {\bibinfo {volume} {81}},\ \bibinfo {pages} {104019} (\bibinfo {year} {2010})},\ \Eprint {http://arxiv.org/abs/0912.5297} {arXiv:0912.5297 [astro-ph.CO]} \BibitemShut {NoStop}%
\bibitem [{\citenamefont {Carr}\ \emph {et~al.}(2020)\citenamefont {Carr}, \citenamefont {Kohri}, \citenamefont {Sendouda},\ and\ \citenamefont {Yokoyama}}]{Carr:2020gox}%
  \BibitemOpen
  \bibfield  {author} {\bibinfo {author} {\bibfnamefont {B.}~\bibnamefont {Carr}}, \bibinfo {author} {\bibfnamefont {K.}~\bibnamefont {Kohri}}, \bibinfo {author} {\bibfnamefont {Y.}~\bibnamefont {Sendouda}}, \ and\ \bibinfo {author} {\bibfnamefont {J.}~\bibnamefont {Yokoyama}},\ }\href@noop {} {\bibfield  {journal} {\bibinfo  {journal} {arXiv}\ } (\bibinfo {year} {2020})},\ \bibinfo {note} {arXiv:2002.12778},\ \Eprint {http://arxiv.org/abs/2002.12778} {arXiv:2002.12778 [astro-ph.CO]} \BibitemShut {NoStop}%
\bibitem [{\citenamefont {Carr}\ and\ \citenamefont {Kuhnel}(2020)}]{Carr:2020xqk}%
  \BibitemOpen
  \bibfield  {author} {\bibinfo {author} {\bibfnamefont {B.}~\bibnamefont {Carr}}\ and\ \bibinfo {author} {\bibfnamefont {F.}~\bibnamefont {Kuhnel}},\ }\href@noop {} {\bibfield  {journal} {\bibinfo  {journal} {Ann. Rev. Nucl. Part. Sci.}\ ,\ \bibinfo {pages} {170:14.1}} (\bibinfo {year} {2020})},\ \Eprint {http://arxiv.org/abs/2006.02838} {arXiv:2006.02838 [astro-ph.CO]} \BibitemShut {NoStop}%
\bibitem [{\citenamefont {Green}\ and\ \citenamefont {Kavanagh}(2021)}]{Green:2020jor}%
  \BibitemOpen
  \bibfield  {author} {\bibinfo {author} {\bibfnamefont {A.~M.}\ \bibnamefont {Green}}\ and\ \bibinfo {author} {\bibfnamefont {B.~J.}\ \bibnamefont {Kavanagh}},\ }\href {\doibase 10.1088/1361-6471/abc534} {\bibfield  {journal} {\bibinfo  {journal} {J. Phys. G}\ }\textbf {\bibinfo {volume} {48}},\ \bibinfo {pages} {043001} (\bibinfo {year} {2021})},\ \Eprint {http://arxiv.org/abs/2007.10722} {arXiv:2007.10722 [astro-ph.CO]} \BibitemShut {NoStop}%
\bibitem [{\citenamefont {Clesse}\ and\ \citenamefont {Garc{\'\i}a-Bellido}(2018)}]{clesse2018seven}%
  \BibitemOpen
  \bibfield  {author} {\bibinfo {author} {\bibfnamefont {S.}~\bibnamefont {Clesse}}\ and\ \bibinfo {author} {\bibfnamefont {J.}~\bibnamefont {Garc{\'\i}a-Bellido}},\ }\href@noop {} {\bibfield  {journal} {\bibinfo  {journal} {Physics of the Dark Universe}\ }\textbf {\bibinfo {volume} {22}},\ \bibinfo {pages} {137} (\bibinfo {year} {2018})}\BibitemShut {NoStop}%
\bibitem [{\citenamefont {Carr}\ \emph {et~al.}(2021)\citenamefont {Carr}, \citenamefont {Clesse}, \citenamefont {Garc\'\i{}a-Bellido},\ and\ \citenamefont {K\"uhnel}}]{Carr:2019kxo}%
  \BibitemOpen
  \bibfield  {author} {\bibinfo {author} {\bibfnamefont {B.}~\bibnamefont {Carr}}, \bibinfo {author} {\bibfnamefont {S.}~\bibnamefont {Clesse}}, \bibinfo {author} {\bibfnamefont {J.}~\bibnamefont {Garc\'\i{}a-Bellido}}, \ and\ \bibinfo {author} {\bibfnamefont {F.}~\bibnamefont {K\"uhnel}},\ }\href {\doibase 10.1016/j.dark.2020.100755} {\bibfield  {journal} {\bibinfo  {journal} {Phys. Dark Univ.}\ }\textbf {\bibinfo {volume} {31}},\ \bibinfo {pages} {100755} (\bibinfo {year} {2021})},\ \Eprint {http://arxiv.org/abs/1906.08217} {arXiv:1906.08217 [astro-ph.CO]} \BibitemShut {NoStop}%
\bibitem [{\citenamefont {{Chandrasekhar}}(1931)}]{chandra}%
  \BibitemOpen
  \bibfield  {author} {\bibinfo {author} {\bibfnamefont {S.}~\bibnamefont {{Chandrasekhar}}},\ }\href {\doibase 10.1086/143324} {\bibfield  {journal} {\bibinfo  {journal} {Astrophys. J.}\ }\textbf {\bibinfo {volume} {74}},\ \bibinfo {pages} {81} (\bibinfo {year} {1931})}\BibitemShut {NoStop}%
\bibitem [{\citenamefont {{Rakavy}}\ and\ \citenamefont {{Shaviv}}(1967)}]{Rakavy_Shaviv_67}%
  \BibitemOpen
  \bibfield  {author} {\bibinfo {author} {\bibfnamefont {G.}~\bibnamefont {{Rakavy}}}\ and\ \bibinfo {author} {\bibfnamefont {G.}~\bibnamefont {{Shaviv}}},\ }\href {\doibase 10.1086/149204} {\bibfield  {journal} {\bibinfo  {journal} {Astrophys. J.}\ }\textbf {\bibinfo {volume} {148}},\ \bibinfo {pages} {803} (\bibinfo {year} {1967})}\BibitemShut {NoStop}%
\bibitem [{\citenamefont {{Fraley}}(1968)}]{1968Ap&SS...2...96F}%
  \BibitemOpen
  \bibfield  {author} {\bibinfo {author} {\bibfnamefont {G.~S.}\ \bibnamefont {{Fraley}}},\ }\href {\doibase 10.1007/BF00651498} {\bibfield  {journal} {\bibinfo  {journal} {Astrophys. Space Sci.}\ }\textbf {\bibinfo {volume} {2}},\ \bibinfo {pages} {96} (\bibinfo {year} {1968})}\BibitemShut {NoStop}%
\bibitem [{\citenamefont {Niemeyer}\ and\ \citenamefont {Jedamzik}(1998)}]{Niemeyer:1997mt}%
  \BibitemOpen
  \bibfield  {author} {\bibinfo {author} {\bibfnamefont {J.~C.}\ \bibnamefont {Niemeyer}}\ and\ \bibinfo {author} {\bibfnamefont {K.}~\bibnamefont {Jedamzik}},\ }\href {\doibase 10.1103/PhysRevLett.80.5481} {\bibfield  {journal} {\bibinfo  {journal} {Phys.~Rev.~Lett.}\ }\textbf {\bibinfo {volume} {80}},\ \bibinfo {pages} {5481} (\bibinfo {year} {1998})},\ \Eprint {http://arxiv.org/abs/astro-ph/9709072} {arXiv:astro-ph/9709072} \BibitemShut {NoStop}%
\bibitem [{\citenamefont {Jedamzik}(1997)}]{Jedamzik:1996mr}%
  \BibitemOpen
  \bibfield  {author} {\bibinfo {author} {\bibfnamefont {K.}~\bibnamefont {Jedamzik}},\ }\href {\doibase 10.1103/PhysRevD.55.5871} {\bibfield  {journal} {\bibinfo  {journal} {Phys.~Rev.}\ }\textbf {\bibinfo {volume} {D55}},\ \bibinfo {pages} {R5871} (\bibinfo {year} {1997})}\BibitemShut {NoStop}%
\bibitem [{\citenamefont {Byrnes}\ \emph {et~al.}(2018)\citenamefont {Byrnes}, \citenamefont {Hindmarsh}, \citenamefont {Young},\ and\ \citenamefont {Hawkins}}]{Byrnes:2018clq}%
  \BibitemOpen
  \bibfield  {author} {\bibinfo {author} {\bibfnamefont {C.~T.}\ \bibnamefont {Byrnes}}, \bibinfo {author} {\bibfnamefont {M.}~\bibnamefont {Hindmarsh}}, \bibinfo {author} {\bibfnamefont {S.}~\bibnamefont {Young}}, \ and\ \bibinfo {author} {\bibfnamefont {M.~R.~S.}\ \bibnamefont {Hawkins}},\ }\href {\doibase 10.1088/1475-7516/2018/08/041} {\bibfield  {journal} {\bibinfo  {journal} {JCAP}\ }\textbf {\bibinfo {volume} {1808}},\ \bibinfo {pages} {041} (\bibinfo {year} {2018})},\ \Eprint {http://arxiv.org/abs/1801.06138} {arXiv:1801.06138 [astro-ph.CO]} \BibitemShut {NoStop}%
\bibitem [{\citenamefont {Clesse}\ and\ \citenamefont {Garcia-Bellido}(2022)}]{Clesse:2020ghq}%
  \BibitemOpen
  \bibfield  {author} {\bibinfo {author} {\bibfnamefont {S.}~\bibnamefont {Clesse}}\ and\ \bibinfo {author} {\bibfnamefont {J.}~\bibnamefont {Garcia-Bellido}},\ }\href {\doibase https://doi.org/10.1016/j.dark.2022.101111} {\bibfield  {journal} {\bibinfo  {journal} {Phys. Dark Univ.}\ }\textbf {\bibinfo {volume} {38}},\ \bibinfo {pages} {101111} (\bibinfo {year} {2022})},\ \Eprint {http://arxiv.org/abs/2007.06481} {arXiv:2007.06481 [astro-ph.CO]} \BibitemShut {NoStop}%
\bibitem [{\citenamefont {Jedamzik}(2021)}]{Jedamzik:2020omx}%
  \BibitemOpen
  \bibfield  {author} {\bibinfo {author} {\bibfnamefont {K.}~\bibnamefont {Jedamzik}},\ }\href {\doibase 10.1103/PhysRevLett.126.051302} {\bibfield  {journal} {\bibinfo  {journal} {Phys. Rev. Lett.}\ }\textbf {\bibinfo {volume} {126}},\ \bibinfo {pages} {051302} (\bibinfo {year} {2021})},\ \Eprint {http://arxiv.org/abs/2007.03565} {arXiv:2007.03565 [astro-ph.CO]} \BibitemShut {NoStop}%
\bibitem [{\citenamefont {Jedamzik}(2020)}]{Jedamzik:2020ypm}%
  \BibitemOpen
  \bibfield  {author} {\bibinfo {author} {\bibfnamefont {K.}~\bibnamefont {Jedamzik}},\ }\href {\doibase 10.1088/1475-7516/2020/09/022} {\bibfield  {journal} {\bibinfo  {journal} {JCAP}\ }\textbf {\bibinfo {volume} {09}},\ \bibinfo {pages} {022} (\bibinfo {year} {2020})},\ \Eprint {http://arxiv.org/abs/2006.11172} {arXiv:2006.11172 [astro-ph.CO]} \BibitemShut {NoStop}%
\bibitem [{\citenamefont {{Wyrzykowski}}\ and\ \citenamefont {{Mandel}}(2020)}]{2020A&A...636A..20W}%
  \BibitemOpen
  \bibfield  {author} {\bibinfo {author} {\bibfnamefont {{\L}.}~\bibnamefont {{Wyrzykowski}}}\ and\ \bibinfo {author} {\bibfnamefont {I.}~\bibnamefont {{Mandel}}},\ }\href {\doibase 10.1051/0004-6361/201935842} {\bibfield  {journal} {\bibinfo  {journal} {Astron. Astrophys.}\ }\textbf {\bibinfo {volume} {636}},\ \bibinfo {eid} {A20} (\bibinfo {year} {2020})},\ \Eprint {http://arxiv.org/abs/1904.07789} {arXiv:1904.07789 [astro-ph.SR]} \BibitemShut {NoStop}%
\bibitem [{\citenamefont {Kashlinsky}(2016)}]{Kashlinsky:2016sdv}%
  \BibitemOpen
  \bibfield  {author} {\bibinfo {author} {\bibfnamefont {A.}~\bibnamefont {Kashlinsky}},\ }\href {\doibase 10.3847/2041-8205/823/2/L25} {\bibfield  {journal} {\bibinfo  {journal} {Astrophys. J. Lett.}\ }\textbf {\bibinfo {volume} {823}},\ \bibinfo {pages} {L25} (\bibinfo {year} {2016})},\ \Eprint {http://arxiv.org/abs/1605.04023} {arXiv:1605.04023 [astro-ph.CO]} \BibitemShut {NoStop}%
\bibitem [{\citenamefont {{Kashlinsky}}\ \emph {et~al.}(2018)\citenamefont {{Kashlinsky}}, \citenamefont {{Arendt}}, \citenamefont {{Atrio-Barandela}}, \citenamefont {{Cappelluti}}, \citenamefont {{Ferrara}},\ and\ \citenamefont {{Hasinger}}}]{2018RvMP...90b5006K}%
  \BibitemOpen
  \bibfield  {author} {\bibinfo {author} {\bibfnamefont {A.}~\bibnamefont {{Kashlinsky}}}, \bibinfo {author} {\bibfnamefont {R.~G.}\ \bibnamefont {{Arendt}}}, \bibinfo {author} {\bibfnamefont {F.}~\bibnamefont {{Atrio-Barandela}}}, \bibinfo {author} {\bibfnamefont {N.}~\bibnamefont {{Cappelluti}}}, \bibinfo {author} {\bibfnamefont {A.}~\bibnamefont {{Ferrara}}}, \ and\ \bibinfo {author} {\bibfnamefont {G.}~\bibnamefont {{Hasinger}}},\ }\href {\doibase 10.1103/RevModPhys.90.025006} {\bibfield  {journal} {\bibinfo  {journal} {Reviews of Modern Physics}\ }\textbf {\bibinfo {volume} {90}},\ \bibinfo {eid} {025006} (\bibinfo {year} {2018})},\ \Eprint {http://arxiv.org/abs/1802.07774} {arXiv:1802.07774 [astro-ph.CO]} \BibitemShut {NoStop}%
\bibitem [{\citenamefont {{Kashlinsky}}\ \emph {et~al.}(2005)\citenamefont {{Kashlinsky}}, \citenamefont {{Arendt}}, \citenamefont {{Mather}},\ and\ \citenamefont {{Moseley}}}]{2005Natur.438...45K}%
  \BibitemOpen
  \bibfield  {author} {\bibinfo {author} {\bibfnamefont {A.}~\bibnamefont {{Kashlinsky}}}, \bibinfo {author} {\bibfnamefont {R.~G.}\ \bibnamefont {{Arendt}}}, \bibinfo {author} {\bibfnamefont {J.}~\bibnamefont {{Mather}}}, \ and\ \bibinfo {author} {\bibfnamefont {S.~H.}\ \bibnamefont {{Moseley}}},\ }\href {\doibase 10.1038/nature04143} {\bibfield  {journal} {\bibinfo  {journal} {\nat}\ }\textbf {\bibinfo {volume} {438}},\ \bibinfo {pages} {45} (\bibinfo {year} {2005})},\ \Eprint {http://arxiv.org/abs/astro-ph/0511105} {arXiv:astro-ph/0511105 [astro-ph]} \BibitemShut {NoStop}%
\bibitem [{\citenamefont {Abbott}\ \emph {et~al.}(2018{\natexlab{a}})\citenamefont {Abbott} \emph {et~al.}}]{Abbott:2018oah}%
  \BibitemOpen
  \bibfield  {author} {\bibinfo {author} {\bibfnamefont {B.~P.}\ \bibnamefont {Abbott}} \emph {et~al.} (\bibinfo {collaboration} {LIGO Scientific, Virgo}),\ }\href {\doibase 10.1103/PhysRevLett.121.231103} {\bibfield  {journal} {\bibinfo  {journal} {Phys. Rev. Lett.}\ }\textbf {\bibinfo {volume} {121}},\ \bibinfo {pages} {231103} (\bibinfo {year} {2018}{\natexlab{a}})},\ \Eprint {http://arxiv.org/abs/1808.04771} {arXiv:1808.04771 [astro-ph.CO]} \BibitemShut {NoStop}%
\bibitem [{\citenamefont {Abbott}\ \emph {et~al.}(2019{\natexlab{b}})\citenamefont {Abbott} \emph {et~al.}}]{LIGOScientific:2019kan}%
  \BibitemOpen
  \bibfield  {author} {\bibinfo {author} {\bibfnamefont {B.~P.}\ \bibnamefont {Abbott}} \emph {et~al.} (\bibinfo {collaboration} {LIGO Scientific, Virgo}),\ }\href {\doibase 10.1103/PhysRevLett.123.161102} {\bibfield  {journal} {\bibinfo  {journal} {Phys. Rev. Lett.}\ }\textbf {\bibinfo {volume} {123}},\ \bibinfo {pages} {161102} (\bibinfo {year} {2019}{\natexlab{b}})},\ \Eprint {http://arxiv.org/abs/1904.08976} {arXiv:1904.08976 [astro-ph.CO]} \BibitemShut {NoStop}%
\bibitem [{\citenamefont {Nitz}\ and\ \citenamefont {Wang}(2021{\natexlab{a}})}]{Nitz:2020bdb}%
  \BibitemOpen
  \bibfield  {author} {\bibinfo {author} {\bibfnamefont {A.~H.}\ \bibnamefont {Nitz}}\ and\ \bibinfo {author} {\bibfnamefont {Y.-F.}\ \bibnamefont {Wang}},\ }\href {\doibase 10.1103/PhysRevLett.126.021103} {\bibfield  {journal} {\bibinfo  {journal} {Phys. Rev. Lett.}\ }\textbf {\bibinfo {volume} {126}},\ \bibinfo {pages} {021103} (\bibinfo {year} {2021}{\natexlab{a}})},\ \Eprint {http://arxiv.org/abs/2007.03583} {arXiv:2007.03583 [astro-ph.HE]} \BibitemShut {NoStop}%
\bibitem [{\citenamefont {Nitz}\ and\ \citenamefont {Wang}(2021{\natexlab{b}})}]{Nitz:2021mzz}%
  \BibitemOpen
  \bibfield  {author} {\bibinfo {author} {\bibfnamefont {A.~H.}\ \bibnamefont {Nitz}}\ and\ \bibinfo {author} {\bibfnamefont {Y.-F.}\ \bibnamefont {Wang}},\ }\href@noop {} {\bibfield  {journal} {\bibinfo  {journal} {arXiv}\ } (\bibinfo {year} {2021}{\natexlab{b}})},\ \Eprint {http://arxiv.org/abs/2102.00868} {arXiv:2102.00868 [astro-ph.HE]} \BibitemShut {NoStop}%
\bibitem [{\citenamefont {Abbott}\ \emph {et~al.}(2022)\citenamefont {Abbott} \emph {et~al.}}]{LIGOScientific:2021job}%
  \BibitemOpen
  \bibfield  {author} {\bibinfo {author} {\bibfnamefont {R.}~\bibnamefont {Abbott}} \emph {et~al.} (\bibinfo {collaboration} {LIGO Scientific, VIRGO, KAGRA}),\ }\href {\doibase 10.1103/PhysRevLett.129.061104} {\bibfield  {journal} {\bibinfo  {journal} {Phys. Rev. Lett.}\ }\textbf {\bibinfo {volume} {129}},\ \bibinfo {pages} {061104} (\bibinfo {year} {2022})},\ \Eprint {http://arxiv.org/abs/2109.12197} {arXiv:2109.12197 [astro-ph.CO]} \BibitemShut {NoStop}%
\bibitem [{\citenamefont {Abbott}\ \emph {et~al.}(2023)\citenamefont {Abbott} \emph {et~al.}}]{LIGOScientific:2022hai}%
  \BibitemOpen
  \bibfield  {author} {\bibinfo {author} {\bibfnamefont {R.}~\bibnamefont {Abbott}} \emph {et~al.} (\bibinfo {collaboration} {LVK}),\ }\href {\doibase 10.1093/mnras/stad588} {\bibfield  {journal} {\bibinfo  {journal} {Mon. Not. Roy. Astron. Soc.}\ }\textbf {\bibinfo {volume} {524}},\ \bibinfo {pages} {5984} (\bibinfo {year} {2023})},\ \bibinfo {note} {[Erratum: Mon.Not.Roy.Astron.Soc. 526, 6234 (2023)]},\ \Eprint {http://arxiv.org/abs/2212.01477} {arXiv:2212.01477 [astro-ph.HE]} \BibitemShut {NoStop}%
\bibitem [{\citenamefont {Abac}\ \emph {et~al.}(2026{\natexlab{c}})\citenamefont {Abac} \emph {et~al.}}]{LIGOScientific:2026wxz}%
  \BibitemOpen
  \bibfield  {author} {\bibinfo {author} {\bibfnamefont {A.~G.}\ \bibnamefont {Abac}} \emph {et~al.} (\bibinfo {collaboration} {LIGO Scientific, VIRGO, KAGRA}),\ }\href@noop {} {\enquote {\bibinfo {title} {{Searches for Binary Mergers with Sub-solar Mass Components in Data from the First Part of LIGO--Virgo--KAGRA's Fourth Observing Run}},}\ } (\bibinfo {year} {2026}{\natexlab{c}}),\ \Eprint {http://arxiv.org/abs/2605.05444} {arXiv:2605.05444 [astro-ph.HE]} \BibitemShut {NoStop}%
\bibitem [{\citenamefont {Raidal}\ \emph {et~al.}(2019)\citenamefont {Raidal}, \citenamefont {Spethmann}, \citenamefont {Vaskonen},\ and\ \citenamefont {Veerm\"ae}}]{Raidal:2018bbj}%
  \BibitemOpen
  \bibfield  {author} {\bibinfo {author} {\bibfnamefont {M.}~\bibnamefont {Raidal}}, \bibinfo {author} {\bibfnamefont {C.}~\bibnamefont {Spethmann}}, \bibinfo {author} {\bibfnamefont {V.}~\bibnamefont {Vaskonen}}, \ and\ \bibinfo {author} {\bibfnamefont {H.}~\bibnamefont {Veerm\"ae}},\ }\href {\doibase 10.1088/1475-7516/2019/02/018} {\bibfield  {journal} {\bibinfo  {journal} {JCAP}\ }\textbf {\bibinfo {volume} {02}},\ \bibinfo {pages} {018} (\bibinfo {year} {2019})},\ \Eprint {http://arxiv.org/abs/1812.01930} {arXiv:1812.01930 [astro-ph.CO]} \BibitemShut {NoStop}%
\bibitem [{\citenamefont {Cannon}\ \emph {et~al.}(2021)\citenamefont {Cannon} \emph {et~al.}}]{Cannon:2020qnf}%
  \BibitemOpen
  \bibfield  {author} {\bibinfo {author} {\bibfnamefont {K.}~\bibnamefont {Cannon}} \emph {et~al.},\ }\href {\doibase https://doi.org/10.1016/j.softx.2021.100680} {\bibfield  {journal} {\bibinfo  {journal} {SoftwareX}\ }\textbf {\bibinfo {volume} {14}},\ \bibinfo {pages} {100680} (\bibinfo {year} {2021})},\ \Eprint {http://arxiv.org/abs/2010.05082} {arXiv:2010.05082 [astro-ph.IM]} \BibitemShut {NoStop}%
\bibitem [{\citenamefont {Cannon}\ \emph {et~al.}(2012)\citenamefont {Cannon}, \citenamefont {Cariou}, \citenamefont {Chapman}, \citenamefont {Crispin-Ortuzar}, \citenamefont {Fotopoulos}, \citenamefont {Frei}, \citenamefont {Hanna}, \citenamefont {Kara}, \citenamefont {Keppel}, \citenamefont {Liao} \emph {et~al.}}]{cannon2012toward}%
  \BibitemOpen
  \bibfield  {author} {\bibinfo {author} {\bibfnamefont {K.}~\bibnamefont {Cannon}}, \bibinfo {author} {\bibfnamefont {R.}~\bibnamefont {Cariou}}, \bibinfo {author} {\bibfnamefont {A.}~\bibnamefont {Chapman}}, \bibinfo {author} {\bibfnamefont {M.}~\bibnamefont {Crispin-Ortuzar}}, \bibinfo {author} {\bibfnamefont {N.}~\bibnamefont {Fotopoulos}}, \bibinfo {author} {\bibfnamefont {M.}~\bibnamefont {Frei}}, \bibinfo {author} {\bibfnamefont {C.}~\bibnamefont {Hanna}}, \bibinfo {author} {\bibfnamefont {E.}~\bibnamefont {Kara}}, \bibinfo {author} {\bibfnamefont {D.}~\bibnamefont {Keppel}}, \bibinfo {author} {\bibfnamefont {L.}~\bibnamefont {Liao}},  \emph {et~al.},\ }\href@noop {} {\bibfield  {journal} {\bibinfo  {journal} {The Astrophysical Journal}\ }\textbf {\bibinfo {volume} {748}},\ \bibinfo {pages} {136} (\bibinfo {year} {2012})}\BibitemShut {NoStop}%
\bibitem [{\citenamefont {Hanna}\ \emph {et~al.}(2020)\citenamefont {Hanna} \emph {et~al.}}]{Hanna:2019ezx}%
  \BibitemOpen
  \bibfield  {author} {\bibinfo {author} {\bibfnamefont {C.}~\bibnamefont {Hanna}} \emph {et~al.},\ }\href {\doibase 10.1103/PhysRevD.101.022003} {\bibfield  {journal} {\bibinfo  {journal} {Phys. Rev. D}\ }\textbf {\bibinfo {volume} {101}},\ \bibinfo {pages} {022003} (\bibinfo {year} {2020})},\ \Eprint {http://arxiv.org/abs/1901.02227} {arXiv:1901.02227 [gr-qc]} \BibitemShut {NoStop}%
\bibitem [{\citenamefont {Messick}\ \emph {et~al.}(2017)\citenamefont {Messick}, \citenamefont {Blackburn}, \citenamefont {Brady}, \citenamefont {Brockill}, \citenamefont {Cannon}, \citenamefont {Cariou}, \citenamefont {Caudill}, \citenamefont {Chamberlin}, \citenamefont {Creighton}, \citenamefont {Everett} \emph {et~al.}}]{messick2017analysis}%
  \BibitemOpen
  \bibfield  {author} {\bibinfo {author} {\bibfnamefont {C.}~\bibnamefont {Messick}}, \bibinfo {author} {\bibfnamefont {K.}~\bibnamefont {Blackburn}}, \bibinfo {author} {\bibfnamefont {P.}~\bibnamefont {Brady}}, \bibinfo {author} {\bibfnamefont {P.}~\bibnamefont {Brockill}}, \bibinfo {author} {\bibfnamefont {K.}~\bibnamefont {Cannon}}, \bibinfo {author} {\bibfnamefont {R.}~\bibnamefont {Cariou}}, \bibinfo {author} {\bibfnamefont {S.}~\bibnamefont {Caudill}}, \bibinfo {author} {\bibfnamefont {S.~J.}\ \bibnamefont {Chamberlin}}, \bibinfo {author} {\bibfnamefont {J.~D.}\ \bibnamefont {Creighton}}, \bibinfo {author} {\bibfnamefont {R.}~\bibnamefont {Everett}},  \emph {et~al.},\ }\href@noop {} {\bibfield  {journal} {\bibinfo  {journal} {Physical Review D}\ }\textbf {\bibinfo {volume} {95}},\ \bibinfo {pages} {042001} (\bibinfo {year} {2017})}\BibitemShut {NoStop}%
\bibitem [{\citenamefont {Sachdev}\ \emph {et~al.}(2019)\citenamefont {Sachdev}, \citenamefont {Caudill}, \citenamefont {Fong}, \citenamefont {Lo}, \citenamefont {Messick}, \citenamefont {Mukherjee}, \citenamefont {Magee}, \citenamefont {Tsukada}, \citenamefont {Blackburn}, \citenamefont {Brady} \emph {et~al.}}]{sachdev2019gstlal}%
  \BibitemOpen
  \bibfield  {author} {\bibinfo {author} {\bibfnamefont {S.}~\bibnamefont {Sachdev}}, \bibinfo {author} {\bibfnamefont {S.}~\bibnamefont {Caudill}}, \bibinfo {author} {\bibfnamefont {H.}~\bibnamefont {Fong}}, \bibinfo {author} {\bibfnamefont {R.~K.}\ \bibnamefont {Lo}}, \bibinfo {author} {\bibfnamefont {C.}~\bibnamefont {Messick}}, \bibinfo {author} {\bibfnamefont {D.}~\bibnamefont {Mukherjee}}, \bibinfo {author} {\bibfnamefont {R.}~\bibnamefont {Magee}}, \bibinfo {author} {\bibfnamefont {L.}~\bibnamefont {Tsukada}}, \bibinfo {author} {\bibfnamefont {K.}~\bibnamefont {Blackburn}}, \bibinfo {author} {\bibfnamefont {P.}~\bibnamefont {Brady}},  \emph {et~al.},\ }\href@noop {} {\bibfield  {journal} {\bibinfo  {journal} {arXiv preprint arXiv:1901.08580}\ } (\bibinfo {year} {2019})}\BibitemShut {NoStop}%
\bibitem [{\citenamefont {{LIGO Scientific Collaboration}}(2018)}]{lalsuite}%
  \BibitemOpen
  \bibfield  {author} {\bibinfo {author} {\bibnamefont {{LIGO Scientific Collaboration}}},\ }\href {\doibase 10.7935/GT1W-FZ16} {\enquote {\bibinfo {title} {{LIGO} {A}lgorithm {L}ibrary - {LALS}uite},}\ }\bibinfo {howpublished} {free software (GPL)} (\bibinfo {year} {2018})\BibitemShut {NoStop}%
\bibitem [{\citenamefont {Dhurandhar}\ and\ \citenamefont {Sathyaprakash}(1994)}]{Dhurandhar:1992mw}%
  \BibitemOpen
  \bibfield  {author} {\bibinfo {author} {\bibfnamefont {S.~V.}\ \bibnamefont {Dhurandhar}}\ and\ \bibinfo {author} {\bibfnamefont {B.~S.}\ \bibnamefont {Sathyaprakash}},\ }\href {\doibase 10.1103/PhysRevD.49.1707} {\bibfield  {journal} {\bibinfo  {journal} {Phys. Rev. D}\ }\textbf {\bibinfo {volume} {49}},\ \bibinfo {pages} {1707} (\bibinfo {year} {1994})}\BibitemShut {NoStop}%
\bibitem [{\citenamefont {Balasubramanian}\ \emph {et~al.}(1996)\citenamefont {Balasubramanian}, \citenamefont {Sathyaprakash},\ and\ \citenamefont {Dhurandhar}}]{Balasubramanian:1995bm}%
  \BibitemOpen
  \bibfield  {author} {\bibinfo {author} {\bibfnamefont {R.}~\bibnamefont {Balasubramanian}}, \bibinfo {author} {\bibfnamefont {B.~S.}\ \bibnamefont {Sathyaprakash}}, \ and\ \bibinfo {author} {\bibfnamefont {S.~V.}\ \bibnamefont {Dhurandhar}},\ }\href {\doibase 10.1103/PhysRevD.53.3033} {\bibfield  {journal} {\bibinfo  {journal} {Phys. Rev. D}\ }\textbf {\bibinfo {volume} {53}},\ \bibinfo {pages} {3033} (\bibinfo {year} {1996})},\ \bibinfo {note} {[Erratum: Phys.Rev.D 54, 1860 (1996)]},\ \Eprint {http://arxiv.org/abs/gr-qc/9508011} {arXiv:gr-qc/9508011} \BibitemShut {NoStop}%
\bibitem [{\citenamefont {Owen}(1996)}]{Owen:1995tm}%
  \BibitemOpen
  \bibfield  {author} {\bibinfo {author} {\bibfnamefont {B.~J.}\ \bibnamefont {Owen}},\ }\href {\doibase 10.1103/PhysRevD.53.6749} {\bibfield  {journal} {\bibinfo  {journal} {Phys. Rev. D}\ }\textbf {\bibinfo {volume} {53}},\ \bibinfo {pages} {6749} (\bibinfo {year} {1996})},\ \Eprint {http://arxiv.org/abs/gr-qc/9511032} {arXiv:gr-qc/9511032} \BibitemShut {NoStop}%
\bibitem [{\citenamefont {Owen}\ and\ \citenamefont {Sathyaprakash}(1999)}]{Owen:1998dk}%
  \BibitemOpen
  \bibfield  {author} {\bibinfo {author} {\bibfnamefont {B.~J.}\ \bibnamefont {Owen}}\ and\ \bibinfo {author} {\bibfnamefont {B.~S.}\ \bibnamefont {Sathyaprakash}},\ }\href {\doibase 10.1103/PhysRevD.60.022002} {\bibfield  {journal} {\bibinfo  {journal} {Phys. Rev. D}\ }\textbf {\bibinfo {volume} {60}},\ \bibinfo {pages} {022002} (\bibinfo {year} {1999})},\ \Eprint {http://arxiv.org/abs/gr-qc/9808076} {arXiv:gr-qc/9808076} \BibitemShut {NoStop}%
\bibitem [{\citenamefont {Allen}\ \emph {et~al.}(2012)\citenamefont {Allen}, \citenamefont {Anderson}, \citenamefont {Brady}, \citenamefont {Brown},\ and\ \citenamefont {Creighton}}]{Allen:2005fk}%
  \BibitemOpen
  \bibfield  {author} {\bibinfo {author} {\bibfnamefont {B.}~\bibnamefont {Allen}}, \bibinfo {author} {\bibfnamefont {W.~G.}\ \bibnamefont {Anderson}}, \bibinfo {author} {\bibfnamefont {P.~R.}\ \bibnamefont {Brady}}, \bibinfo {author} {\bibfnamefont {D.~A.}\ \bibnamefont {Brown}}, \ and\ \bibinfo {author} {\bibfnamefont {J.~D.~E.}\ \bibnamefont {Creighton}},\ }\href {\doibase 10.1103/PhysRevD.85.122006} {\bibfield  {journal} {\bibinfo  {journal} {Phys. Rev. D}\ }\textbf {\bibinfo {volume} {85}},\ \bibinfo {pages} {122006} (\bibinfo {year} {2012})},\ \Eprint {http://arxiv.org/abs/gr-qc/0509116} {arXiv:gr-qc/0509116} \BibitemShut {NoStop}%
\bibitem [{\citenamefont {Blanchet}\ \emph {et~al.}(1995)\citenamefont {Blanchet}, \citenamefont {Damour}, \citenamefont {Iyer}, \citenamefont {Will},\ and\ \citenamefont {Wiseman}}]{Blanchet:1995ez}%
  \BibitemOpen
  \bibfield  {author} {\bibinfo {author} {\bibfnamefont {L.}~\bibnamefont {Blanchet}}, \bibinfo {author} {\bibfnamefont {T.}~\bibnamefont {Damour}}, \bibinfo {author} {\bibfnamefont {B.~R.}\ \bibnamefont {Iyer}}, \bibinfo {author} {\bibfnamefont {C.~M.}\ \bibnamefont {Will}}, \ and\ \bibinfo {author} {\bibfnamefont {A.~G.}\ \bibnamefont {Wiseman}},\ }\href {\doibase 10.1103/PhysRevLett.74.3515} {\bibfield  {journal} {\bibinfo  {journal} {Phys. Rev. Lett.}\ }\textbf {\bibinfo {volume} {74}},\ \bibinfo {pages} {3515} (\bibinfo {year} {1995})},\ \Eprint {http://arxiv.org/abs/gr-qc/9501027} {arXiv:gr-qc/9501027} \BibitemShut {NoStop}%
\bibitem [{\citenamefont {Poisson}(1998)}]{Poisson:1997ha}%
  \BibitemOpen
  \bibfield  {author} {\bibinfo {author} {\bibfnamefont {E.}~\bibnamefont {Poisson}},\ }\href {\doibase 10.1103/PhysRevD.57.5287} {\bibfield  {journal} {\bibinfo  {journal} {Phys. Rev. D}\ }\textbf {\bibinfo {volume} {57}},\ \bibinfo {pages} {5287} (\bibinfo {year} {1998})},\ \Eprint {http://arxiv.org/abs/gr-qc/9709032} {arXiv:gr-qc/9709032} \BibitemShut {NoStop}%
\bibitem [{\citenamefont {Damour}\ \emph {et~al.}(2001{\natexlab{a}})\citenamefont {Damour}, \citenamefont {Jaranowski},\ and\ \citenamefont {Schaefer}}]{Damour:2001bu}%
  \BibitemOpen
  \bibfield  {author} {\bibinfo {author} {\bibfnamefont {T.}~\bibnamefont {Damour}}, \bibinfo {author} {\bibfnamefont {P.}~\bibnamefont {Jaranowski}}, \ and\ \bibinfo {author} {\bibfnamefont {G.}~\bibnamefont {Schaefer}},\ }\href {\doibase 10.1016/S0370-2693(01)00642-6} {\bibfield  {journal} {\bibinfo  {journal} {Phys. Lett. B}\ }\textbf {\bibinfo {volume} {513}},\ \bibinfo {pages} {147} (\bibinfo {year} {2001}{\natexlab{a}})},\ \Eprint {http://arxiv.org/abs/gr-qc/0105038} {arXiv:gr-qc/0105038} \BibitemShut {NoStop}%
\bibitem [{\citenamefont {Damour}\ \emph {et~al.}(2001{\natexlab{b}})\citenamefont {Damour}, \citenamefont {Iyer},\ and\ \citenamefont {Sathyaprakash}}]{Damour:2000zb}%
  \BibitemOpen
  \bibfield  {author} {\bibinfo {author} {\bibfnamefont {T.}~\bibnamefont {Damour}}, \bibinfo {author} {\bibfnamefont {B.~R.}\ \bibnamefont {Iyer}}, \ and\ \bibinfo {author} {\bibfnamefont {B.~S.}\ \bibnamefont {Sathyaprakash}},\ }\href {\doibase 10.1103/PhysRevD.63.044023} {\bibfield  {journal} {\bibinfo  {journal} {Phys. Rev. D}\ }\textbf {\bibinfo {volume} {63}},\ \bibinfo {pages} {044023} (\bibinfo {year} {2001}{\natexlab{b}})},\ \bibinfo {note} {[Erratum: Phys.Rev.D 72, 029902 (2005)]},\ \Eprint {http://arxiv.org/abs/gr-qc/0010009} {arXiv:gr-qc/0010009} \BibitemShut {NoStop}%
\bibitem [{\citenamefont {Buonanno}\ \emph {et~al.}(2009)\citenamefont {Buonanno}, \citenamefont {Iyer}, \citenamefont {Ochsner}, \citenamefont {Pan},\ and\ \citenamefont {Sathyaprakash}}]{Buonanno:2009zt}%
  \BibitemOpen
  \bibfield  {author} {\bibinfo {author} {\bibfnamefont {A.}~\bibnamefont {Buonanno}}, \bibinfo {author} {\bibfnamefont {B.}~\bibnamefont {Iyer}}, \bibinfo {author} {\bibfnamefont {E.}~\bibnamefont {Ochsner}}, \bibinfo {author} {\bibfnamefont {Y.}~\bibnamefont {Pan}}, \ and\ \bibinfo {author} {\bibfnamefont {B.~S.}\ \bibnamefont {Sathyaprakash}},\ }\href {\doibase 10.1103/PhysRevD.80.084043} {\bibfield  {journal} {\bibinfo  {journal} {Phys. Rev. D}\ }\textbf {\bibinfo {volume} {80}},\ \bibinfo {pages} {084043} (\bibinfo {year} {2009})},\ \Eprint {http://arxiv.org/abs/0907.0700} {arXiv:0907.0700 [gr-qc]} \BibitemShut {NoStop}%
\bibitem [{\citenamefont {Mikoczi}\ \emph {et~al.}(2005)\citenamefont {Mikoczi}, \citenamefont {Vasuth},\ and\ \citenamefont {Gergely}}]{Mikoczi:2005dn}%
  \BibitemOpen
  \bibfield  {author} {\bibinfo {author} {\bibfnamefont {B.}~\bibnamefont {Mikoczi}}, \bibinfo {author} {\bibfnamefont {M.}~\bibnamefont {Vasuth}}, \ and\ \bibinfo {author} {\bibfnamefont {L.~A.}\ \bibnamefont {Gergely}},\ }\href {\doibase 10.1103/PhysRevD.71.124043} {\bibfield  {journal} {\bibinfo  {journal} {Phys. Rev. D}\ }\textbf {\bibinfo {volume} {71}},\ \bibinfo {pages} {124043} (\bibinfo {year} {2005})},\ \Eprint {http://arxiv.org/abs/astro-ph/0504538} {arXiv:astro-ph/0504538} \BibitemShut {NoStop}%
\bibitem [{\citenamefont {Blanchet}\ \emph {et~al.}(2005)\citenamefont {Blanchet}, \citenamefont {Damour}, \citenamefont {Esposito-Farese},\ and\ \citenamefont {Iyer}}]{Blanchet:2005tk}%
  \BibitemOpen
  \bibfield  {author} {\bibinfo {author} {\bibfnamefont {L.}~\bibnamefont {Blanchet}}, \bibinfo {author} {\bibfnamefont {T.}~\bibnamefont {Damour}}, \bibinfo {author} {\bibfnamefont {G.}~\bibnamefont {Esposito-Farese}}, \ and\ \bibinfo {author} {\bibfnamefont {B.~R.}\ \bibnamefont {Iyer}},\ }\href {\doibase 10.1103/PhysRevD.71.124004} {\bibfield  {journal} {\bibinfo  {journal} {Phys. Rev. D}\ }\textbf {\bibinfo {volume} {71}},\ \bibinfo {pages} {124004} (\bibinfo {year} {2005})},\ \Eprint {http://arxiv.org/abs/gr-qc/0503044} {arXiv:gr-qc/0503044} \BibitemShut {NoStop}%
\bibitem [{\citenamefont {Arun}\ \emph {et~al.}(2009)\citenamefont {Arun}, \citenamefont {Buonanno}, \citenamefont {Faye},\ and\ \citenamefont {Ochsner}}]{Arun:2008kb}%
  \BibitemOpen
  \bibfield  {author} {\bibinfo {author} {\bibfnamefont {K.~G.}\ \bibnamefont {Arun}}, \bibinfo {author} {\bibfnamefont {A.}~\bibnamefont {Buonanno}}, \bibinfo {author} {\bibfnamefont {G.}~\bibnamefont {Faye}}, \ and\ \bibinfo {author} {\bibfnamefont {E.}~\bibnamefont {Ochsner}},\ }\href {\doibase 10.1103/PhysRevD.79.104023} {\bibfield  {journal} {\bibinfo  {journal} {Phys. Rev. D}\ }\textbf {\bibinfo {volume} {79}},\ \bibinfo {pages} {104023} (\bibinfo {year} {2009})},\ \bibinfo {note} {[Erratum: Phys.Rev.D 84, 049901 (2011)]},\ \Eprint {http://arxiv.org/abs/0810.5336} {arXiv:0810.5336 [gr-qc]} \BibitemShut {NoStop}%
\bibitem [{\citenamefont {Boh\'e}\ \emph {et~al.}(2013)\citenamefont {Boh\'e}, \citenamefont {Marsat},\ and\ \citenamefont {Blanchet}}]{Bohe:2013cla}%
  \BibitemOpen
  \bibfield  {author} {\bibinfo {author} {\bibfnamefont {A.}~\bibnamefont {Boh\'e}}, \bibinfo {author} {\bibfnamefont {S.}~\bibnamefont {Marsat}}, \ and\ \bibinfo {author} {\bibfnamefont {L.}~\bibnamefont {Blanchet}},\ }\href {\doibase 10.1088/0264-9381/30/13/135009} {\bibfield  {journal} {\bibinfo  {journal} {Class. Quant. Grav.}\ }\textbf {\bibinfo {volume} {30}},\ \bibinfo {pages} {135009} (\bibinfo {year} {2013})},\ \Eprint {http://arxiv.org/abs/1303.7412} {arXiv:1303.7412 [gr-qc]} \BibitemShut {NoStop}%
\bibitem [{\citenamefont {Boh\'e}\ \emph {et~al.}(2015)\citenamefont {Boh\'e}, \citenamefont {Faye}, \citenamefont {Marsat},\ and\ \citenamefont {Porter}}]{Bohe:2015ana}%
  \BibitemOpen
  \bibfield  {author} {\bibinfo {author} {\bibfnamefont {A.}~\bibnamefont {Boh\'e}}, \bibinfo {author} {\bibfnamefont {G.}~\bibnamefont {Faye}}, \bibinfo {author} {\bibfnamefont {S.}~\bibnamefont {Marsat}}, \ and\ \bibinfo {author} {\bibfnamefont {E.~K.}\ \bibnamefont {Porter}},\ }\href {\doibase 10.1088/0264-9381/32/19/195010} {\bibfield  {journal} {\bibinfo  {journal} {Class. Quant. Grav.}\ }\textbf {\bibinfo {volume} {32}},\ \bibinfo {pages} {195010} (\bibinfo {year} {2015})},\ \Eprint {http://arxiv.org/abs/1501.01529} {arXiv:1501.01529 [gr-qc]} \BibitemShut {NoStop}%
\bibitem [{\citenamefont {Mishra}\ \emph {et~al.}(2016)\citenamefont {Mishra}, \citenamefont {Kela}, \citenamefont {Arun},\ and\ \citenamefont {Faye}}]{Mishra:2016whh}%
  \BibitemOpen
  \bibfield  {author} {\bibinfo {author} {\bibfnamefont {C.~K.}\ \bibnamefont {Mishra}}, \bibinfo {author} {\bibfnamefont {A.}~\bibnamefont {Kela}}, \bibinfo {author} {\bibfnamefont {K.~G.}\ \bibnamefont {Arun}}, \ and\ \bibinfo {author} {\bibfnamefont {G.}~\bibnamefont {Faye}},\ }\href {\doibase 10.1103/PhysRevD.93.084054} {\bibfield  {journal} {\bibinfo  {journal} {Phys. Rev. D}\ }\textbf {\bibinfo {volume} {93}},\ \bibinfo {pages} {084054} (\bibinfo {year} {2016})},\ \Eprint {http://arxiv.org/abs/1601.05588} {arXiv:1601.05588 [gr-qc]} \BibitemShut {NoStop}%
\bibitem [{\citenamefont {Harry}\ \emph {et~al.}(2008)\citenamefont {Harry}, \citenamefont {Fairhurst},\ and\ \citenamefont {Sathyaprakash}}]{Harry:2008yn}%
  \BibitemOpen
  \bibfield  {author} {\bibinfo {author} {\bibfnamefont {I.~W.}\ \bibnamefont {Harry}}, \bibinfo {author} {\bibfnamefont {S.}~\bibnamefont {Fairhurst}}, \ and\ \bibinfo {author} {\bibfnamefont {B.~S.}\ \bibnamefont {Sathyaprakash}},\ }\href {\doibase 10.1088/0264-9381/25/18/184027} {\bibfield  {journal} {\bibinfo  {journal} {Class. Quant. Grav.}\ }\textbf {\bibinfo {volume} {25}},\ \bibinfo {pages} {184027} (\bibinfo {year} {2008})},\ \Eprint {http://arxiv.org/abs/0804.3274} {arXiv:0804.3274 [gr-qc]} \BibitemShut {NoStop}%
\bibitem [{\citenamefont {Babak}(2008)}]{Babak:2008rb}%
  \BibitemOpen
  \bibfield  {author} {\bibinfo {author} {\bibfnamefont {S.}~\bibnamefont {Babak}},\ }\href {\doibase 10.1088/0264-9381/25/19/195011} {\bibfield  {journal} {\bibinfo  {journal} {Class. Quant. Grav.}\ }\textbf {\bibinfo {volume} {25}},\ \bibinfo {pages} {195011} (\bibinfo {year} {2008})},\ \Eprint {http://arxiv.org/abs/0801.4070} {arXiv:0801.4070 [gr-qc]} \BibitemShut {NoStop}%
\bibitem [{\citenamefont {Harry}\ \emph {et~al.}(2009)\citenamefont {Harry}, \citenamefont {Allen},\ and\ \citenamefont {Sathyaprakash}}]{Harry:2009ea}%
  \BibitemOpen
  \bibfield  {author} {\bibinfo {author} {\bibfnamefont {I.~W.}\ \bibnamefont {Harry}}, \bibinfo {author} {\bibfnamefont {B.}~\bibnamefont {Allen}}, \ and\ \bibinfo {author} {\bibfnamefont {B.~S.}\ \bibnamefont {Sathyaprakash}},\ }\href {\doibase 10.1103/PhysRevD.80.104014} {\bibfield  {journal} {\bibinfo  {journal} {Phys. Rev. D}\ }\textbf {\bibinfo {volume} {80}},\ \bibinfo {pages} {104014} (\bibinfo {year} {2009})},\ \Eprint {http://arxiv.org/abs/0908.2090} {arXiv:0908.2090 [gr-qc]} \BibitemShut {NoStop}%
\bibitem [{\citenamefont {Manca}\ and\ \citenamefont {Vallisneri}(2010)}]{Manca:2009xw}%
  \BibitemOpen
  \bibfield  {author} {\bibinfo {author} {\bibfnamefont {G.~M.}\ \bibnamefont {Manca}}\ and\ \bibinfo {author} {\bibfnamefont {M.}~\bibnamefont {Vallisneri}},\ }\href {\doibase 10.1103/PhysRevD.81.024004} {\bibfield  {journal} {\bibinfo  {journal} {Phys. Rev. D}\ }\textbf {\bibinfo {volume} {81}},\ \bibinfo {pages} {024004} (\bibinfo {year} {2010})},\ \Eprint {http://arxiv.org/abs/0909.0563} {arXiv:0909.0563 [gr-qc]} \BibitemShut {NoStop}%
\bibitem [{\citenamefont {Magee}\ \emph {et~al.}(2018)\citenamefont {Magee}, \citenamefont {Deutsch}, \citenamefont {McClincy}, \citenamefont {Hanna}, \citenamefont {Horst}, \citenamefont {Meacher}, \citenamefont {Messick}, \citenamefont {Shandera},\ and\ \citenamefont {Wade}}]{Magee:2018opb}%
  \BibitemOpen
  \bibfield  {author} {\bibinfo {author} {\bibfnamefont {R.}~\bibnamefont {Magee}}, \bibinfo {author} {\bibfnamefont {A.-S.}\ \bibnamefont {Deutsch}}, \bibinfo {author} {\bibfnamefont {P.}~\bibnamefont {McClincy}}, \bibinfo {author} {\bibfnamefont {C.}~\bibnamefont {Hanna}}, \bibinfo {author} {\bibfnamefont {C.}~\bibnamefont {Horst}}, \bibinfo {author} {\bibfnamefont {D.}~\bibnamefont {Meacher}}, \bibinfo {author} {\bibfnamefont {C.}~\bibnamefont {Messick}}, \bibinfo {author} {\bibfnamefont {S.}~\bibnamefont {Shandera}}, \ and\ \bibinfo {author} {\bibfnamefont {M.}~\bibnamefont {Wade}},\ }\href {\doibase 10.1103/PhysRevD.98.103024} {\bibfield  {journal} {\bibinfo  {journal} {Phys. Rev. D}\ }\textbf {\bibinfo {volume} {98}},\ \bibinfo {pages} {103024} (\bibinfo {year} {2018})},\ \Eprint {http://arxiv.org/abs/1808.04772} {arXiv:1808.04772 [astro-ph.IM]} \BibitemShut {NoStop}%
\bibitem [{\citenamefont {Cannon}\ \emph {et~al.}(2015)\citenamefont {Cannon}, \citenamefont {Hanna},\ and\ \citenamefont {Peoples}}]{Cannon:2015gha}%
  \BibitemOpen
  \bibfield  {author} {\bibinfo {author} {\bibfnamefont {K.}~\bibnamefont {Cannon}}, \bibinfo {author} {\bibfnamefont {C.}~\bibnamefont {Hanna}}, \ and\ \bibinfo {author} {\bibfnamefont {J.}~\bibnamefont {Peoples}},\ }\href@noop {} {\enquote {\bibinfo {title} {{Likelihood-Ratio Ranking Statistic for Compact Binary Coalescence Candidates with Rate Estimation}},}\ } (\bibinfo {year} {2015}),\ \Eprint {http://arxiv.org/abs/1504.04632} {arXiv:1504.04632 [astro-ph.IM]} \BibitemShut {NoStop}%
\bibitem [{\citenamefont {Fong}(2018)}]{Fong2018Thesis}%
  \BibitemOpen
  \bibfield  {author} {\bibinfo {author} {\bibfnamefont {H.~K.~Y.}\ \bibnamefont {Fong}},\ }\emph {\bibinfo {title} {From simulations to signals: Analyzing gravitational waves from compact binary coalescences}},\ \href {http://hdl.handle.net/1807/91831} {Ph.D. thesis},\ \bibinfo  {school} {University of Toronto} (\bibinfo {year} {2018})\BibitemShut {NoStop}%
\bibitem [{\citenamefont {Essick}\ \emph {et~al.}(2020)\citenamefont {Essick}, \citenamefont {Godwin}, \citenamefont {Hanna}, \citenamefont {Blackburn},\ and\ \citenamefont {Katsavounidis}}]{Essick:2020qpo}%
  \BibitemOpen
  \bibfield  {author} {\bibinfo {author} {\bibfnamefont {R.}~\bibnamefont {Essick}}, \bibinfo {author} {\bibfnamefont {P.}~\bibnamefont {Godwin}}, \bibinfo {author} {\bibfnamefont {C.}~\bibnamefont {Hanna}}, \bibinfo {author} {\bibfnamefont {L.}~\bibnamefont {Blackburn}}, \ and\ \bibinfo {author} {\bibfnamefont {E.}~\bibnamefont {Katsavounidis}},\ }\href {\doibase 10.1088/2632-2153/abab5f} {\bibfield  {journal} {\bibinfo  {journal} {Machine Learning: Science and Technology}\ }\textbf {\bibinfo {volume} {2}},\ \bibinfo {pages} {015004} (\bibinfo {year} {2020})},\ \Eprint {http://arxiv.org/abs/2005.12761} {arXiv:2005.12761 [astro-ph.IM]} \BibitemShut {NoStop}%
\bibitem [{\citenamefont {Godwin}\ \emph {et~al.}(2020)\citenamefont {Godwin} \emph {et~al.}}]{Godwin:2020weu}%
  \BibitemOpen
  \bibfield  {author} {\bibinfo {author} {\bibfnamefont {P.}~\bibnamefont {Godwin}} \emph {et~al.},\ }\href@noop {} {\enquote {\bibinfo {title} {{Incorporation of Statistical Data Quality Information into the GstLAL Search Analysis}},}\ } (\bibinfo {year} {2020}),\ \Eprint {http://arxiv.org/abs/2010.15282} {arXiv:2010.15282 [gr-qc]} \BibitemShut {NoStop}%
\bibitem [{\citenamefont {Cannon}\ \emph {et~al.}(2013)\citenamefont {Cannon}, \citenamefont {Hanna},\ and\ \citenamefont {Keppel}}]{Cannon:2012zt}%
  \BibitemOpen
  \bibfield  {author} {\bibinfo {author} {\bibfnamefont {K.}~\bibnamefont {Cannon}}, \bibinfo {author} {\bibfnamefont {C.}~\bibnamefont {Hanna}}, \ and\ \bibinfo {author} {\bibfnamefont {D.}~\bibnamefont {Keppel}},\ }\href {\doibase 10.1103/PhysRevD.88.024025} {\bibfield  {journal} {\bibinfo  {journal} {Phys. Rev. D}\ }\textbf {\bibinfo {volume} {88}},\ \bibinfo {pages} {024025} (\bibinfo {year} {2013})},\ \Eprint {http://arxiv.org/abs/1209.0718} {arXiv:1209.0718 [gr-qc]} \BibitemShut {NoStop}%
\bibitem [{\citenamefont {Morras}\ \emph {et~al.}(2023)\citenamefont {Morras} \emph {et~al.}}]{Morras:2023jvb}%
  \BibitemOpen
  \bibfield  {author} {\bibinfo {author} {\bibfnamefont {G.}~\bibnamefont {Morras}} \emph {et~al.},\ }\href {\doibase 10.1016/j.dark.2023.101285} {\bibfield  {journal} {\bibinfo  {journal} {Phys. Dark Univ.}\ }\textbf {\bibinfo {volume} {42}},\ \bibinfo {pages} {101285} (\bibinfo {year} {2023})},\ \Eprint {http://arxiv.org/abs/2301.11619} {arXiv:2301.11619 [gr-qc]} \BibitemShut {NoStop}%
\bibitem [{\citenamefont {Farr}\ \emph {et~al.}(2015)\citenamefont {Farr}, \citenamefont {Gair}, \citenamefont {Mandel},\ and\ \citenamefont {Cutler}}]{Farr:2013yna}%
  \BibitemOpen
  \bibfield  {author} {\bibinfo {author} {\bibfnamefont {W.~M.}\ \bibnamefont {Farr}}, \bibinfo {author} {\bibfnamefont {J.~R.}\ \bibnamefont {Gair}}, \bibinfo {author} {\bibfnamefont {I.}~\bibnamefont {Mandel}}, \ and\ \bibinfo {author} {\bibfnamefont {C.}~\bibnamefont {Cutler}},\ }\href {\doibase 10.1103/PhysRevD.91.023005} {\bibfield  {journal} {\bibinfo  {journal} {Phys. Rev. D}\ }\textbf {\bibinfo {volume} {91}},\ \bibinfo {pages} {023005} (\bibinfo {year} {2015})},\ \Eprint {http://arxiv.org/abs/1302.5341} {arXiv:1302.5341 [astro-ph.IM]} \BibitemShut {NoStop}%
\bibitem [{\citenamefont {Kapadia}\ \emph {et~al.}(2020)\citenamefont {Kapadia} \emph {et~al.}}]{Kapadia:2019uut}%
  \BibitemOpen
  \bibfield  {author} {\bibinfo {author} {\bibfnamefont {S.~J.}\ \bibnamefont {Kapadia}} \emph {et~al.},\ }\href {\doibase 10.1088/1361-6382/ab5f2d} {\bibfield  {journal} {\bibinfo  {journal} {Class. Quant. Grav.}\ }\textbf {\bibinfo {volume} {37}},\ \bibinfo {pages} {045007} (\bibinfo {year} {2020})},\ \Eprint {http://arxiv.org/abs/1903.06881} {arXiv:1903.06881 [astro-ph.HE]} \BibitemShut {NoStop}%
\bibitem [{\citenamefont {Gorton}\ and\ \citenamefont {Green}(2022)}]{Gorton:2022fyb}%
  \BibitemOpen
  \bibfield  {author} {\bibinfo {author} {\bibfnamefont {M.}~\bibnamefont {Gorton}}\ and\ \bibinfo {author} {\bibfnamefont {A.~M.}\ \bibnamefont {Green}},\ }\href {\doibase 10.1088/1475-7516/2022/08/035} {\bibfield  {journal} {\bibinfo  {journal} {JCAP}\ }\textbf {\bibinfo {volume} {08}},\ \bibinfo {pages} {035} (\bibinfo {year} {2022})},\ \Eprint {http://arxiv.org/abs/2203.04209} {arXiv:2203.04209 [astro-ph.CO]} \BibitemShut {NoStop}%
\bibitem [{\citenamefont {Peta{\v{c}}}\ \emph {et~al.}(2022)\citenamefont {Peta{\v{c}}}, \citenamefont {Lavalle},\ and\ \citenamefont {Jedamzik}}]{Petac:2022rio}%
  \BibitemOpen
  \bibfield  {author} {\bibinfo {author} {\bibfnamefont {M.}~\bibnamefont {Peta{\v{c}}}}, \bibinfo {author} {\bibfnamefont {J.}~\bibnamefont {Lavalle}}, \ and\ \bibinfo {author} {\bibfnamefont {K.}~\bibnamefont {Jedamzik}},\ }\href {\doibase 10.1103/PhysRevD.105.083520} {\bibfield  {journal} {\bibinfo  {journal} {Phys. Rev. D}\ }\textbf {\bibinfo {volume} {105}},\ \bibinfo {pages} {083520} (\bibinfo {year} {2022})},\ \Eprint {http://arxiv.org/abs/2201.02521} {arXiv:2201.02521 [astro-ph.CO]} \BibitemShut {NoStop}%
\bibitem [{\citenamefont {De~Luca}\ \emph {et~al.}(2022)\citenamefont {De~Luca}, \citenamefont {Franciolini}, \citenamefont {Riotto},\ and\ \citenamefont {Veerm{\"a}e}}]{DeLuca:2022uvz}%
  \BibitemOpen
  \bibfield  {author} {\bibinfo {author} {\bibfnamefont {V.}~\bibnamefont {De~Luca}}, \bibinfo {author} {\bibfnamefont {G.}~\bibnamefont {Franciolini}}, \bibinfo {author} {\bibfnamefont {A.}~\bibnamefont {Riotto}}, \ and\ \bibinfo {author} {\bibfnamefont {H.}~\bibnamefont {Veerm{\"a}e}},\ }\href {\doibase 10.1103/PhysRevLett.129.191302} {\bibfield  {journal} {\bibinfo  {journal} {Phys. Rev. Lett.}\ }\textbf {\bibinfo {volume} {129}},\ \bibinfo {pages} {191302} (\bibinfo {year} {2022})},\ \Eprint {http://arxiv.org/abs/2208.01683} {arXiv:2208.01683 [astro-ph.CO]} \BibitemShut {NoStop}%
\bibitem [{\citenamefont {Abbott}\ \emph {et~al.}(2020{\natexlab{e}})\citenamefont {Abbott} \emph {et~al.}}]{Abbott:2020uma}%
  \BibitemOpen
  \bibfield  {author} {\bibinfo {author} {\bibfnamefont {B.~P.}\ \bibnamefont {Abbott}} \emph {et~al.} (\bibinfo {collaboration} {LIGO Scientific, Virgo}),\ }\href {\doibase 10.3847/2041-8213/ab75f5} {\bibfield  {journal} {\bibinfo  {journal} {Astrophys. J. Lett.}\ }\textbf {\bibinfo {volume} {892}},\ \bibinfo {pages} {L3} (\bibinfo {year} {2020}{\natexlab{e}})},\ \Eprint {http://arxiv.org/abs/2001.01761} {arXiv:2001.01761 [astro-ph.HE]} \BibitemShut {NoStop}%
\bibitem [{\citenamefont {Ali-Ha\"\i{}moud}\ \emph {et~al.}(2017)\citenamefont {Ali-Ha\"\i{}moud}, \citenamefont {Kovetz},\ and\ \citenamefont {Kamionkowski}}]{Ali-Haimoud:2017rtz}%
  \BibitemOpen
  \bibfield  {author} {\bibinfo {author} {\bibfnamefont {Y.}~\bibnamefont {Ali-Ha\"\i{}moud}}, \bibinfo {author} {\bibfnamefont {E.~D.}\ \bibnamefont {Kovetz}}, \ and\ \bibinfo {author} {\bibfnamefont {M.}~\bibnamefont {Kamionkowski}},\ }\href {\doibase 10.1103/PhysRevD.96.123523} {\bibfield  {journal} {\bibinfo  {journal} {Phys. Rev. D}\ }\textbf {\bibinfo {volume} {96}},\ \bibinfo {pages} {123523} (\bibinfo {year} {2017})},\ \Eprint {http://arxiv.org/abs/1709.06576} {arXiv:1709.06576 [astro-ph.CO]} \BibitemShut {NoStop}%
\bibitem [{\citenamefont {Gow}\ \emph {et~al.}(2020)\citenamefont {Gow}, \citenamefont {Byrnes}, \citenamefont {Hall},\ and\ \citenamefont {Peacock}}]{Gow:2019pok}%
  \BibitemOpen
  \bibfield  {author} {\bibinfo {author} {\bibfnamefont {A.~D.}\ \bibnamefont {Gow}}, \bibinfo {author} {\bibfnamefont {C.~T.}\ \bibnamefont {Byrnes}}, \bibinfo {author} {\bibfnamefont {A.}~\bibnamefont {Hall}}, \ and\ \bibinfo {author} {\bibfnamefont {J.~A.}\ \bibnamefont {Peacock}},\ }\href {\doibase 10.1088/1475-7516/2020/01/031} {\bibfield  {journal} {\bibinfo  {journal} {JCAP}\ }\textbf {\bibinfo {volume} {01}},\ \bibinfo {pages} {031} (\bibinfo {year} {2020})},\ \Eprint {http://arxiv.org/abs/1911.12685} {arXiv:1911.12685 [astro-ph.CO]} \BibitemShut {NoStop}%
\bibitem [{\citenamefont {Liu}\ \emph {et~al.}(2019)\citenamefont {Liu}, \citenamefont {Guo},\ and\ \citenamefont {Cai}}]{Liu:2019rnx}%
  \BibitemOpen
  \bibfield  {author} {\bibinfo {author} {\bibfnamefont {L.}~\bibnamefont {Liu}}, \bibinfo {author} {\bibfnamefont {Z.-K.}\ \bibnamefont {Guo}}, \ and\ \bibinfo {author} {\bibfnamefont {R.-G.}\ \bibnamefont {Cai}},\ }\href {\doibase 10.1140/epjc/s10052-019-7227-0} {\bibfield  {journal} {\bibinfo  {journal} {Eur. Phys. J. C}\ }\textbf {\bibinfo {volume} {79}},\ \bibinfo {pages} {717} (\bibinfo {year} {2019})},\ \Eprint {http://arxiv.org/abs/1901.07672} {arXiv:1901.07672 [astro-ph.CO]} \BibitemShut {NoStop}%
\bibitem [{\citenamefont {Kocsis}\ \emph {et~al.}(2018)\citenamefont {Kocsis}, \citenamefont {Suyama}, \citenamefont {Tanaka},\ and\ \citenamefont {Yokoyama}}]{Kocsis_2018}%
  \BibitemOpen
  \bibfield  {author} {\bibinfo {author} {\bibfnamefont {B.}~\bibnamefont {Kocsis}}, \bibinfo {author} {\bibfnamefont {T.}~\bibnamefont {Suyama}}, \bibinfo {author} {\bibfnamefont {T.}~\bibnamefont {Tanaka}}, \ and\ \bibinfo {author} {\bibfnamefont {S.}~\bibnamefont {Yokoyama}},\ }\href {\doibase 10.3847/1538-4357/aaa7f4} {\bibfield  {journal} {\bibinfo  {journal} {The Astrophysical Journal}\ }\textbf {\bibinfo {volume} {854}},\ \bibinfo {pages} {41} (\bibinfo {year} {2018})}\BibitemShut {NoStop}%
\bibitem [{\citenamefont {H\"utsi}\ \emph {et~al.}(2021{\natexlab{a}})\citenamefont {H\"utsi}, \citenamefont {Raidal}, \citenamefont {Vaskonen},\ and\ \citenamefont {Veerm\"ae}}]{Hutsi:2020sol}%
  \BibitemOpen
  \bibfield  {author} {\bibinfo {author} {\bibfnamefont {G.}~\bibnamefont {H\"utsi}}, \bibinfo {author} {\bibfnamefont {M.}~\bibnamefont {Raidal}}, \bibinfo {author} {\bibfnamefont {V.}~\bibnamefont {Vaskonen}}, \ and\ \bibinfo {author} {\bibfnamefont {H.}~\bibnamefont {Veerm\"ae}},\ }\href {\doibase 10.1088/1475-7516/2021/03/068} {\bibfield  {journal} {\bibinfo  {journal} {JCAP}\ }\textbf {\bibinfo {volume} {03}},\ \bibinfo {pages} {068} (\bibinfo {year} {2021}{\natexlab{a}})},\ \Eprint {http://arxiv.org/abs/2012.02786} {arXiv:2012.02786 [astro-ph.CO]} \BibitemShut {NoStop}%
\bibitem [{\citenamefont {Vaskonen}\ and\ \citenamefont {Veerm\"ae}(2020)}]{Vaskonen:2019jpv}%
  \BibitemOpen
  \bibfield  {author} {\bibinfo {author} {\bibfnamefont {V.}~\bibnamefont {Vaskonen}}\ and\ \bibinfo {author} {\bibfnamefont {H.}~\bibnamefont {Veerm\"ae}},\ }\href {\doibase 10.1103/PhysRevD.101.043015} {\bibfield  {journal} {\bibinfo  {journal} {Phys. Rev. D}\ }\textbf {\bibinfo {volume} {101}},\ \bibinfo {pages} {043015} (\bibinfo {year} {2020})},\ \Eprint {http://arxiv.org/abs/1908.09752} {arXiv:1908.09752 [astro-ph.CO]} \BibitemShut {NoStop}%
\bibitem [{\citenamefont {Abbott}\ \emph {et~al.}(2020{\natexlab{f}})\citenamefont {Abbott} \emph {et~al.}}]{LIGOScientific:2020aai}%
  \BibitemOpen
  \bibfield  {author} {\bibinfo {author} {\bibfnamefont {B.~P.}\ \bibnamefont {Abbott}} \emph {et~al.} (\bibinfo {collaboration} {LIGO Scientific, Virgo}),\ }\href {\doibase 10.3847/2041-8213/ab75f5} {\bibfield  {journal} {\bibinfo  {journal} {Astrophys. J. Lett.}\ }\textbf {\bibinfo {volume} {892}},\ \bibinfo {pages} {L3} (\bibinfo {year} {2020}{\natexlab{f}})},\ \Eprint {http://arxiv.org/abs/2001.01761} {arXiv:2001.01761 [astro-ph.HE]} \BibitemShut {NoStop}%
\bibitem [{\citenamefont {Abbott}\ \emph {et~al.}(2020{\natexlab{g}})\citenamefont {Abbott} \emph {et~al.}}]{LIGOScientific:2020zkf}%
  \BibitemOpen
  \bibfield  {author} {\bibinfo {author} {\bibfnamefont {R.}~\bibnamefont {Abbott}} \emph {et~al.} (\bibinfo {collaboration} {LIGO Scientific, Virgo}),\ }\href {\doibase 10.3847/2041-8213/ab960f} {\bibfield  {journal} {\bibinfo  {journal} {Astrophys. J. Lett.}\ }\textbf {\bibinfo {volume} {896}},\ \bibinfo {pages} {L44} (\bibinfo {year} {2020}{\natexlab{g}})},\ \Eprint {http://arxiv.org/abs/2006.12611} {arXiv:2006.12611 [astro-ph.HE]} \BibitemShut {NoStop}%
\bibitem [{\citenamefont {Abbott}\ \emph {et~al.}(2020{\natexlab{h}})\citenamefont {Abbott} \emph {et~al.}}]{LIGOScientific:2020iuh}%
  \BibitemOpen
  \bibfield  {author} {\bibinfo {author} {\bibfnamefont {R.}~\bibnamefont {Abbott}} \emph {et~al.} (\bibinfo {collaboration} {LIGO Scientific, Virgo}),\ }\href {\doibase 10.1103/PhysRevLett.125.101102} {\bibfield  {journal} {\bibinfo  {journal} {Phys. Rev. Lett.}\ }\textbf {\bibinfo {volume} {125}},\ \bibinfo {pages} {101102} (\bibinfo {year} {2020}{\natexlab{h}})},\ \Eprint {http://arxiv.org/abs/2009.01075} {arXiv:2009.01075 [gr-qc]} \BibitemShut {NoStop}%
\bibitem [{\citenamefont {Raidal}\ \emph {et~al.}(2025)\citenamefont {Raidal}, \citenamefont {Vaskonen},\ and\ \citenamefont {Veerm{\"a}e}}]{Raidal:2024bmm}%
  \BibitemOpen
  \bibfield  {author} {\bibinfo {author} {\bibfnamefont {M.}~\bibnamefont {Raidal}}, \bibinfo {author} {\bibfnamefont {V.}~\bibnamefont {Vaskonen}}, \ and\ \bibinfo {author} {\bibfnamefont {H.}~\bibnamefont {Veerm{\"a}e}},\ }\enquote {\bibinfo {title} {{Formation of~Primordial Black Hole Binaries and~Their Merger Rates}},}\ in\ \href {\doibase 10.1007/978-981-97-8887-3_16} {\emph {\bibinfo {booktitle} {{Primordial Black Holes}}}},\ \bibinfo {editor} {edited by\ \bibinfo {editor} {\bibfnamefont {C.}~\bibnamefont {Byrnes}}, \bibinfo {editor} {\bibfnamefont {G.}~\bibnamefont {Franciolini}}, \bibinfo {editor} {\bibfnamefont {T.}~\bibnamefont {Harada}}, \bibinfo {editor} {\bibfnamefont {P.}~\bibnamefont {Pani}}, \ and\ \bibinfo {editor} {\bibfnamefont {M.}~\bibnamefont {Sasaki}}}\ (\bibinfo {year} {2025})\ \Eprint {http://arxiv.org/abs/2404.08416} {arXiv:2404.08416 [astro-ph.CO]} \BibitemShut {NoStop}%
\bibitem [{\citenamefont {Buikema}\ \emph {et~al.}(2020)\citenamefont {Buikema} \emph {et~al.}}]{aLIGO:2020wna}%
  \BibitemOpen
  \bibfield  {author} {\bibinfo {author} {\bibfnamefont {A.}~\bibnamefont {Buikema}} \emph {et~al.} (\bibinfo {collaboration} {aLIGO}),\ }\href {\doibase 10.1103/PhysRevD.102.062003} {\bibfield  {journal} {\bibinfo  {journal} {Phys. Rev. D}\ }\textbf {\bibinfo {volume} {102}},\ \bibinfo {pages} {062003} (\bibinfo {year} {2020})},\ \Eprint {http://arxiv.org/abs/2008.01301} {arXiv:2008.01301 [astro-ph.IM]} \BibitemShut {NoStop}%
\bibitem [{\citenamefont {Acernese}\ \emph {et~al.}(2023)\citenamefont {Acernese} \emph {et~al.}}]{Virgo:2022ysc}%
  \BibitemOpen
  \bibfield  {author} {\bibinfo {author} {\bibfnamefont {F.}~\bibnamefont {Acernese}} \emph {et~al.} (\bibinfo {collaboration} {Virgo}),\ }\href {\doibase 10.1088/1361-6382/acd92d} {\bibfield  {journal} {\bibinfo  {journal} {Class. Quant. Grav.}\ }\textbf {\bibinfo {volume} {40}},\ \bibinfo {pages} {185006} (\bibinfo {year} {2023})},\ \Eprint {http://arxiv.org/abs/2210.15633} {arXiv:2210.15633 [gr-qc]} \BibitemShut {NoStop}%
\bibitem [{\citenamefont {Capote}\ \emph {et~al.}(2025)\citenamefont {Capote} \emph {et~al.}}]{Capote:2024rmo}%
  \BibitemOpen
  \bibfield  {author} {\bibinfo {author} {\bibfnamefont {E.}~\bibnamefont {Capote}} \emph {et~al.},\ }\href {\doibase 10.1103/PhysRevD.111.062002} {\bibfield  {journal} {\bibinfo  {journal} {Phys. Rev. D}\ }\textbf {\bibinfo {volume} {111}},\ \bibinfo {pages} {062002} (\bibinfo {year} {2025})},\ \Eprint {http://arxiv.org/abs/2411.14607} {arXiv:2411.14607 [gr-qc]} \BibitemShut {NoStop}%
\bibitem [{\citenamefont {Acernese}\ \emph {et~al.}(2026)\citenamefont {Acernese} \emph {et~al.}}]{VIRGO:2026okv}%
  \BibitemOpen
  \bibfield  {author} {\bibinfo {author} {\bibfnamefont {F.}~\bibnamefont {Acernese}} \emph {et~al.} (\bibinfo {collaboration} {VIRGO}),\ }\href@noop {} {\enquote {\bibinfo {title} {{Advanced Virgo during the LIGO-Virgo-KAGRA fourth observing run}},}\ } (\bibinfo {year} {2026}),\ \Eprint {http://arxiv.org/abs/2607.26872} {arXiv:2607.26872 [gr-qc]} \BibitemShut {NoStop}%
\bibitem [{\citenamefont {De~Luca}\ \emph {et~al.}(2021)\citenamefont {De~Luca}, \citenamefont {Franciolini},\ and\ \citenamefont {Riotto}}]{DeLucae:2020agl}%
  \BibitemOpen
  \bibfield  {author} {\bibinfo {author} {\bibfnamefont {V.}~\bibnamefont {De~Luca}}, \bibinfo {author} {\bibfnamefont {G.}~\bibnamefont {Franciolini}}, \ and\ \bibinfo {author} {\bibfnamefont {A.}~\bibnamefont {Riotto}},\ }\href {\doibase 10.1103/PhysRevLett.126.041303} {\bibfield  {journal} {\bibinfo  {journal} {Phys. Rev. Lett.}\ }\textbf {\bibinfo {volume} {126}},\ \bibinfo {pages} {041303} (\bibinfo {year} {2021})},\ \Eprint {http://arxiv.org/abs/2009.08268} {arXiv:2009.08268 [astro-ph.CO]} \BibitemShut {NoStop}%
\bibitem [{\citenamefont {Boehm}\ \emph {et~al.}(2021)\citenamefont {Boehm}, \citenamefont {Kobakhidze}, \citenamefont {O'hare}, \citenamefont {Picker},\ and\ \citenamefont {Sakellariadou}}]{Boehm:2020jwd}%
  \BibitemOpen
  \bibfield  {author} {\bibinfo {author} {\bibfnamefont {C.}~\bibnamefont {Boehm}}, \bibinfo {author} {\bibfnamefont {A.}~\bibnamefont {Kobakhidze}}, \bibinfo {author} {\bibfnamefont {C.~A.~J.}\ \bibnamefont {O'hare}}, \bibinfo {author} {\bibfnamefont {Z.~S.~C.}\ \bibnamefont {Picker}}, \ and\ \bibinfo {author} {\bibfnamefont {M.}~\bibnamefont {Sakellariadou}},\ }\href {\doibase 10.1088/1475-7516/2021/03/078} {\bibfield  {journal} {\bibinfo  {journal} {JCAP}\ }\textbf {\bibinfo {volume} {03}},\ \bibinfo {pages} {078} (\bibinfo {year} {2021})},\ \Eprint {http://arxiv.org/abs/2008.10743} {arXiv:2008.10743 [astro-ph.CO]} \BibitemShut {NoStop}%
\bibitem [{\citenamefont {De~Luca}\ \emph {et~al.}(2020)\citenamefont {De~Luca}, \citenamefont {Desjacques}, \citenamefont {Franciolini},\ and\ \citenamefont {Riotto}}]{DeLuca:2020jug}%
  \BibitemOpen
  \bibfield  {author} {\bibinfo {author} {\bibfnamefont {V.}~\bibnamefont {De~Luca}}, \bibinfo {author} {\bibfnamefont {V.}~\bibnamefont {Desjacques}}, \bibinfo {author} {\bibfnamefont {G.}~\bibnamefont {Franciolini}}, \ and\ \bibinfo {author} {\bibfnamefont {A.}~\bibnamefont {Riotto}},\ }\href {\doibase 10.1088/1475-7516/2020/11/028} {\bibfield  {journal} {\bibinfo  {journal} {JCAP}\ }\textbf {\bibinfo {volume} {11}},\ \bibinfo {pages} {028} (\bibinfo {year} {2020})},\ \Eprint {http://arxiv.org/abs/2009.04731} {arXiv:2009.04731 [astro-ph.CO]} \BibitemShut {NoStop}%
\bibitem [{\citenamefont {H\"utsi}\ \emph {et~al.}(2021{\natexlab{b}})\citenamefont {H\"utsi}, \citenamefont {Koivisto}, \citenamefont {Raidal}, \citenamefont {Vaskonen},\ and\ \citenamefont {Veerm\"ae}}]{Hutsi:2021nvs}%
  \BibitemOpen
  \bibfield  {author} {\bibinfo {author} {\bibfnamefont {G.}~\bibnamefont {H\"utsi}}, \bibinfo {author} {\bibfnamefont {T.}~\bibnamefont {Koivisto}}, \bibinfo {author} {\bibfnamefont {M.}~\bibnamefont {Raidal}}, \bibinfo {author} {\bibfnamefont {V.}~\bibnamefont {Vaskonen}}, \ and\ \bibinfo {author} {\bibfnamefont {H.}~\bibnamefont {Veerm\"ae}},\ }\href {\doibase 10.1140/epjc/s10052-021-09803-4} {\bibfield  {journal} {\bibinfo  {journal} {Eur. Phys. J. C}\ }\textbf {\bibinfo {volume} {81}},\ \bibinfo {pages} {999} (\bibinfo {year} {2021}{\natexlab{b}})},\ \Eprint {http://arxiv.org/abs/2105.09328} {arXiv:2105.09328 [astro-ph.CO]} \BibitemShut {NoStop}%
\bibitem [{\citenamefont {Abbott}\ \emph {et~al.}(2018{\natexlab{b}})\citenamefont {Abbott}, \citenamefont {Abbott}, \citenamefont {Abbott}, \citenamefont {Abernathy}, \citenamefont {Acernese}, \citenamefont {Ackley}, \citenamefont {Adams}, \citenamefont {Adams}, \citenamefont {Addesso}, \citenamefont {Adhikari} \emph {et~al.}}]{abbott2018effects}%
  \BibitemOpen
  \bibfield  {author} {\bibinfo {author} {\bibfnamefont {B.~P.}\ \bibnamefont {Abbott}}, \bibinfo {author} {\bibfnamefont {R.}~\bibnamefont {Abbott}}, \bibinfo {author} {\bibfnamefont {T.~D.}\ \bibnamefont {Abbott}}, \bibinfo {author} {\bibfnamefont {M.}~\bibnamefont {Abernathy}}, \bibinfo {author} {\bibfnamefont {F.}~\bibnamefont {Acernese}}, \bibinfo {author} {\bibfnamefont {K.}~\bibnamefont {Ackley}}, \bibinfo {author} {\bibfnamefont {C.}~\bibnamefont {Adams}}, \bibinfo {author} {\bibfnamefont {T.}~\bibnamefont {Adams}}, \bibinfo {author} {\bibfnamefont {P.}~\bibnamefont {Addesso}}, \bibinfo {author} {\bibfnamefont {R.}~\bibnamefont {Adhikari}},  \emph {et~al.},\ }\href@noop {} {\bibfield  {journal} {\bibinfo  {journal} {Classical and Quantum Gravity}\ }\textbf {\bibinfo {volume} {35}},\ \bibinfo {pages} {065010} (\bibinfo {year} {2018}{\natexlab{b}})}\BibitemShut {NoStop}%
\bibitem [{\citenamefont {Nitz}\ \emph {et~al.}(2021)\citenamefont {Nitz} \emph {et~al.}}]{alex_nitz_2021_4556907}%
  \BibitemOpen
  \bibfield  {author} {\bibinfo {author} {\bibfnamefont {A.}~\bibnamefont {Nitz}} \emph {et~al.},\ }\href {\doibase 10.5281/zenodo.4556907} {\enquote {\bibinfo {title} {gwastro/pycbc: 1.18.0 release of pycbc},}\ } (\bibinfo {year} {2021})\BibitemShut {NoStop}%
\end{thebibliography}%

\end{document}